\documentstyle{article}

\begin{document}

\def \Pi {P}

\title{Geometrodynamics of Schwarzschild Black Holes} \author{Karel
V. Kucha\v{r} \\ Department of Physics \\ University of Utah \\ Salt
Lake City, Utah 84112, U.S.A.} \maketitle

\baselineskip=16pt

\begin{abstract}
The curvature coordinates $T,R$ of a Schwarz\-schild spacetime are
turned into canonical coordinates $T(r), {\sf R}(r)$ on the phase
space of spherically symmetric black holes. The entire dynamical
content of the Hamiltonian theory is reduced to the constraints
requiring that the momenta $P_{T}(r), P_{\sf R}(r)$ vanish. What
remains is a conjugate pair of canonical variables $m$ and $p$ whose
values are the same on every embedding. The coordinate $m$ is the
Schwarzschild mass, and the momentum $p$ the difference of
parametrization times at right and left infinities. The Dirac
constraint quantization in the new representation leads to the state
functional $\Psi (m; T, {\sf R}] = \Psi (m)$ which describes an
unchanging superposition of black holes with different masses. The
new canonical variables may be employed in the study of collapsing
matter systems.
\end{abstract}

\section{Introduction}

\subsection{Context}

     General relativity was cast into canonical form by Dirac
\cite{Dirac2}, and by Arnowitt, Deser and Misner (ADM) \cite{ADM}.
Inside asymptotic regions, Hamiltonian dynamics is entirely
generated by constraints. Their imposition as operator restrictions
on the states yields the Wheeler-DeWitt equation \cite{Wheeler1},
\cite{DeWitt}.  DeWitt realized that by freezing all but few degrees
of freedom of a cosmological model by symmetry, one can obtain
exactly soluble models of quantum gravity \cite{DeWitt}.  Misner and
his school turned this idea of {\em minisuperspace quantization}
\cite{Misner-mini} into a systematic exploration of quantum
cosmology \cite{Ryan}, \cite{McCallum}.

     Minisuperspace techniques were extended to {\em midisuperspace
quantization} of infinitely-dimensional models.  The first system
treated in this manner was the cylindrical gravitational wave
\cite{Kuchar-cyl}. It became clear that the Wheeler-DeWitt equation
is often unwieldy and difficult to interpret.  For cylindrical
waves, the switch to an extrinsic time representation changed the
Wheeler-DeWitt equation into a functional time Schr\"{o}dinger
equation.  Gravity assumed the form of a parametrized field theory
\cite{Dirac2}, \cite{Kuchar-par}.

     Things did not work that way for other infinitely-dimensional
systems. The most important of these is the gravitational collapse
of a spherically symmetric distribution of matter.  Berger, Chitre,
Moncrief and Nutku (BCMN) set this problem in the Dirac-ADM
midisuperspace formalism. In their classic paper \cite{Berger} they
studied a spherically symmetric massless scalar field coupled to
gravity. They did not succeed in finding an extrinsic time
representation.  Instead, they reduced the action to a privileged
foliation characterized by the vanishing `radial' momentum. Their
reduced Hamiltonian did not quite reproduce the field equations.
This was found and corrected by Unruh \cite{Unruh}.

     The BCMN model opens a canonical route to the study of
Hawking's radiation \cite{Hawking1}, \cite{Hawking2}. A standard
semiclassical analysis of the Hawking effect starts with a black
hole being formed by the gravitational collapse of classical matter.
One studies a field that propagates on this background. The modes
that disappear below the horizon are averaged out, and the thermal
radiation escaping to infinity is described by a density operator.
The Hawking radiation leads to the evaporation of the black hole. No
general agreement has been reached on what is the final state of
this process. The black hole may evaporate completely, or leave a
remnant. If it evaporates completely, the question remains what
happens to the information which got initially trapped below its
horizon.

     Midisuperspace canonical approach has two potential advantages
over the standard description. First, it goes beyond the
semiclassical approximation. Second, unless one encounters a Cauchy
horizon, all information is registered in the canonical data on a
Cauchy hypersurface. One can study what happens inside black holes
and how they approach the singularity. However, to make use of this
advantage, one must insist that the foliation covers all available
spacetime, and that time evolution is not artificially arrested. The
BCMN slicing does not meet this condition.  A careful study of the
BCMN model in its relation to Hawking's radiation was undertaken by
H\'{a}j\'{\i}\v{c}ek \cite{Hajicek1} - \cite{Hajicek3}. He
generalized the model to other spherically symmetric fields, and
paid special attention to the properties of apparent horizon and
choice of slicing.

     One can easily see that the BCMN slicing is problematic already
for primordial Schwarzschild black holes. In vacuo, the BCMN slices
coincide with those of constant Killing time $T$. They cover only
the static regions of the Kruskal diagram, and never penetrate the
horizon. Canonical treatment of a complete Schwarzschild spacetime
was attempted by Lund \cite{Lund}. To get below the horizon, Lund
used the Lamaitre slices, or the slices of constant $R<2M$. He did
not succeed in covering the whole Kruskal diagram by a single
foliation, or relate the state of the Schwarzschild black hole on
$T=const$ slices to its state on the Lamaitre slices or the
$2M>R=const$ slices.  The best solution would have been to work in
the functional time representation, but Lund presented a proof that
the extrinsic time representation does not exist for vacuum
Schwarzschild black holes.

     Interest in primordial black holes has been revived by a surge
of activity on quantization of dilatonic black holes (see recent
reviews by Gidddings \cite{Giddings}, and by Harvey and Strominger
\cite{Strominger}). Starting from this program, Gegenberg,
Kunstatter and Louis-Martinez \cite{Kunstatter1}, \cite{Kunstatter2}
discussed canonical quantization of Schwarzschild black holes within
conformally invariant formulation of Einstein's theory
\cite{Kunstatter3}.  Along different lines, Thiemann and Kastrup
\cite{Thiemann1} - \cite{Thiemann4} discussed canonical quantization
of Schwarzschild and Riesner-Nordstrom black holes mostly, though
not entirely, within Ashtekar's canonical formalism.

     Keeping this background in mind, let us state the goals and
results of this paper.

\subsection{Results}

     We cast the classical and quantum dynamics of primordial black
holes into geometrically transparent and explicitly soluble form by
choosing a natural canonical chart on the phase space. Our method
amounts to finding a functional extrinsic time representation
analogous to one which exists for cylindrical gravitational waves.
Lund's no-go theorem is transcended because the representation does
not satisfy its unnecessarily strong premises.

     Geometrodynamics of Schwarzschild black holes is governed by
the Dirac-ADM action restricted to spacetimes with spherical
symmetry. The spatial metric $g_{ab}(x^{c})$ on a symmetric
hypersurface is entirely characterized by two functions,
$\Lambda(r)$ and $R(r)$, of a radial coordinate $r$. In canonical
formalism, these are accompanied by the conjugate momenta
$P_{\Lambda}(r)$ and $\Pi_{R}(r)$. We choose phase space variables
which have an immediate geometric meaning. The hypersurface action
yields the familiar Hamiltonian and momentum constraints equivalent
to those derived by BCMN \cite{Berger}.

     Proper understanding of the canonical formalism requires a
careful handling of boundary conditions. One must specify how fast
the canonical data and the lapse-shift multipliers fall at
infinities.  The requirement that $\Lambda$ can be freely varied
within its falloff class mandates the addition of the ADM boundary
term to the hypersurface action. Unfortunately, the introduction of
this term prevents one from freely varying the lapse function at
infinities.  This defect is amended by parametrization, which makes
the action dependent on two more variables, namely, on proper times
measured by static clocks at infinities. All the variables in the
parametrized action can be freely varied, and their interconnection
at infinities acquires the status of natural boundary conditions.
This version of the canonical action principle is vital for our
treatment of black hole dynamics.

     The crux of our approach is the introduction of the Killing
time $T(r)$ as a canonical coordinate. The way in which $T(r)$
enters the formalism is rather delicate. To begin with, one
restricts attention to the space of solutions. In a given
Schwarzschild spacetime, one can specify a hypersurface by giving
the familiar curvature coordinates of its points as functions of a
radial label $r$. The Schwarzschild geometry induces on the
hypersurface the canonical data which satisfy the constraints. From
those data, one can locally determine the Schwarzschild mass $M$ of
the embedding spacetime, and the rate $T'(r)$ at which the Killing
time changes along the hypersurface.  Though $T(r)$ becomes infinite
on the horizon, its change across the horizon can be consistently
inferred from smooth canonical data.

     Next, one forgets how the expressions for $M(r)$ and $-T'(r)$
were obtained, and turns them into definitions of two new sets of
dynamical variables on the phase space. At this stage, the canonical
data no longer need to satisfy the constraints, and the mass
function $M(r)$ can in principle depend on $r$. The remarkable
feature of the new variables is that they form a canonical pair:
$P_{M}(r)=-T'(r)$.  Moreover, by retaining the curvature coordinate,
$R(r) \mapsto {\sf R}(r) = R(r)$, but modifying its conjugate
momentum, $P_{R}(r) \mapsto P_{\sf R}(r)$, one can complete $M(r)$
and $-T'(r)$ into a new canonical chart
\begin{equation}
M(r),\, P_{M}(r) = - T'(r)\,;\;{\sf R}(r)=R(r),\, P_{\sf R}(r)\,.
\end{equation}
We exhibit the generating functional of the canonical transformation
from the old to the new canonical variables. Because $T'(r)$ has an
infinite jump on the horizon, the canonical transformation has there
a singularity.

     We can now reimpose the constraints, and evolve the variables
by Hamilton equations. Under these circumstances, the mass function
becomes a position-independent constant of motion.  Inversely, one
can show that the Hamiltonian and momentum constraints can be
replaced by a much simpler set of conditions
\begin{equation}
M'(r)=0\,, \;\;\; P_{{\sf R}}(r)=0\
\end{equation}
on the new canonical variables.

     The canonical structure we have derived holds only when the
lapse function in the ADM boundary action is considered as a fixed
function of the label time. After parametrization, the action
becomes dependent on a pair of proper times $\tau_{\pm}$ at
infinities, and it no longer has a canonical form. It is surprising
that without adding any more variables to the parametrized space
$\tau_{+}, \tau_{-}, M(r), P_{M}(r), {\sf R}(r), P_{\sf R}(r)$, one
can introduce on it a canonical chart
\begin{equation}
m, p;\; T(r), P_{T}(r);\; {\sf R}(r), P_{\sf R}(r) \,.
\label{eq:finalchart}
\end{equation}
The final canonical variables (\ref{eq:finalchart}) have a simple
physical meaning. The canonical pair ${\sf R}(r), P_{\sf R}(r)$
remains unchanged. The new canonical coordinate $T(r)$ is the
Killing time, and its conjugate momentum $P_{T}(r)$ is the mass
density $-M'(r)$. The curvature coordinates $T,R$ in spacetime are
thereby turned into canonical coordinates $T(r), {\sf R}(r)$ on the
phase space.  Simultaneously, the constraints are transformed into
the statement that the momenta $P_{T}(r)$ and $P_{\sf R}(r)$
canonically conjugate to the embedding variables $T(r)$ and ${\sf
R}(r)$ vanish:
\begin{equation}
P_{T}(r) = 0 \,, \;\;\; P_{\sf R}(r) = 0 \,.
\end{equation}
The Hamiltonian is a linear combination of these constraints. The
true ADM Hamiltonian disappeared in the transformation process to
the final canonical chart (\ref{eq:finalchart}). This phenomenon can
be understood as a result of time-dependent canonical
transformation.

     Because the Hamiltonian weakly vanishes, the remaining
canonical variables $m$ and $p$ are constants of motion. The
canonical coordinate $m$ is the Schwarzschild mass of the black
hole. The meaning of $p$ is more esoteric: $p$ characterizes the
difference between the parametrization times at the left and right
infinities.  The comparison of these two times is made possible by
connecting the infinities by hypersurfaces of constant Killing time.
Once set, this difference is preserved because the two
parametrization clocks run at the same rate. This explains why $p$
is a constant of motion. While it has always been suspected that the
variable conjugate to the Schwarzschild mass is some sort of proper
time, the correct interpretation of this quantity eluded previous
investigators, likely because they did not pay enough attention to
the role of parametrization, and to how the two asymptotic regions
are connected with each other.

     The parametrized action can be reduced to true dynamical
degrees of freedom, and the dynamics generated by the unparametrized
action can be compared to that we have just described. It transpires
that our parametrized viewpoint corresponds to performing a
time-dependent canonical transformation to `initial data'.

     With the new polarization (\ref{eq:finalchart}) of the phase
space, the Dirac constraint quantization of primordial black holes
becomes straightforward. The state functional $\Psi (m;\,T, {\sf
R}]$ of the system is reduced by the constraints to an
embedding-independent function $\Psi (m)$ of the mass parameter $m$.
Such a state function describes a superposition of primordial black
holes of different masses. Once prepared, it stays the same on every
hypersurface $T(r), {\sf R}(r)$.

     The curvature coordinates are ill-behaved on the horizon.
However, once they are constructed from the canonical data, one can
easily transform them into Kruskal coordinates by a further
canonical transformation.

     Primordial black holes, despite all the care needed for their
proper canonical treatment, are dynamically trivial. The true
interest of the new canonical variables lies in the possibilities
which they open in the study of gravitational collapse of matter.
These questions are now being pursued in collaboration with Dr.~Petr
H\'{a}j\'{\i}\v{c}ek and Dr.~Joseph Romano.

\section{Schwarzschild Solution}

\subsection{Spacetime Description}

The recurrent theme of this paper is that canonical formalism should
be guided by spacetime intuition. We thus start by summarizing what
is known about spherically symmetric solutions of vacuum Einstein
equations.

     Any such solution is locally isometric to the Schwarzschild
line element
\begin{equation}
ds^{2} = - F(R)\,dT^{2} + F^{-1}(R)\,dR^{2} + R^{2}\,d \Omega^{2}
\label{eq:Schw}
\end{equation}
written in the {\em curvature coordinates} $\big( T, R \big)$. Here,
\begin{equation}
d \Omega^{2} = d \theta ^{2} + \sin ^{2} \theta \, d \phi ^{2}
\end{equation}
is the line element on the unit sphere. We are using natural units
in which the Newton constant of gravitation $G$ and the speed of
light $c$ are put equal to one: $G=1=c$. The coefficient $F(R)$ has
the form
\begin{equation}
F(R) = 1 - 2M/R \,, \label{eq:F}
\end{equation}
where $M$ is a constant. The curvature coordinate $R$ is invariantly
defined by the requirement that $4 \pi R^{2}$ be the area of the
2-spheres S$^{2}$: $\, T= const,\; R=const \,$ which are the
transitivity surfaces of the rotation group. The vector field $
\partial / \partial T$ is a Killing vector field of the metric
(\ref{eq:Schw}). It is orthogonal to the hypersurfaces $T=const$ of
the {\em Killing time} $T$.

     For $M=0$ the spacetime is flat. Solutions with $M>0$ describe
black holes, solutions with $M<0$ correspond to naked singularities.
We limit ourselves to solutions with $M \geq 0$.

     As $R \rightarrow \infty$, the line element (\ref{eq:Schw})
becomes asymptotically flat. At $R=2M$, the solution (\ref{eq:Schw})
runs into a coordinate singularity. The maximal analytic extension
of (\ref{eq:Schw}) from the region $R>2M>0$ across $R=2M$ describes
a primordial black hole. The complete spacetime $\cal M$ is
represented by the familiar Kruskal diagram \cite{Kruskal},
\cite{Hawking-Ellis}.  It is covered by four patches of curvature
coordinates which meet at the horizon $F(R)=0$.  Two regions have
$R>2M$, and the Killing field $\partial / \partial T$ in them is
timelike. We call them the right and left static regions, I and III.
Two other regions have $R<2M$, and the Killing field $\partial /
\partial T$ in them is spacelike. We call them the past and future
dynamical regions, IV and II. The past dynamical region begins and
the future dynamical region ends in a true curvature singularity at
$R=0$.

     The total Schwarzschild spacetime can be covered by a single
patch of {\em Kruskal coordinates} $U$ and $V$. The lines $U =
const$ are radial rightgoing null geodesics $\nearrow$, the lines $V
= const$ are radial leftgoing null geodesics $\nwarrow$. Both $U$
and $V$ grow from past to future.  The horizon is transformed to the
lines $U=0$ and $V=0$. Their intersection is the {\em bifurcation
point}.  The four regions covered by curvature coordinates are:
\begin{description}
\item[right static region I:] $R>2M:\; U<0, V>0\,$,
\item[left static region III:] $R>2M:\;  U>0, V<0\,$,
\item[future dynamical region II:] $R<2M:\; U>0, V>0\,$,
\item[past dynamical region IV:] $R<2M:\; U<0, V<0\,$.
\end{description}

     The Kruskal coordinates $U$ and $V$ are mapped into curvature
coordinates $T$ and $R$ by a two-to-one transformation. Anticipating
the steps we shall need later in the canonical formalism, we build
this transformation in several steps.  The Kruskal coordinates are
dimensionless. We thus first {\em scale} the curvature coordinates
$T$ and $R$ into dimensionless coordinates $\bar{T}$ and $\bar{R}$:
\begin{equation}
\bar{T} = \frac{T}{2M}\,, \;\;\; \bar{R} = \frac{R}{2M}\,.
\label{eq:scale}
\end{equation}
Curvature coordinates are related to Kruskal coordinates by
\begin{equation}
UV = {\cal W}(\bar{R})\;\mbox{$:=$}\; \big( 1 - \bar{R} \big) \,
\exp \big( {\bar{R}} \big) \,, \;\;\; \frac{V}{U} = {\rm sign} \big(
1 - \bar{R} \big) \, \exp \big( {\bar{T}} \big)\,.  \label{eq:UV}
\end{equation}

     The $T=const$ hypersurfaces appear in the Kruskal diagram as
straight lines passing through the bifurcation point. The rightgoing
branch $\nearrow$ of the horizon is labeled by $T=\infty$, the
leftgoing branch $\nwarrow$ by $T=-\infty$. The $R=const$ lines are
hyperbolas asymptotic to the horizon.

     As $\bar{R}$ increases from $\bar{R}=0$ (singularity) to
$\bar{R}=\infty$, the function ${\cal W}(\bar{R})$ monotonically
decreases from $1$ to $-\infty$. Therefore, it has an inverse,
${\cal R}(UV)$. This enables us to solve (\ref{eq:UV}) for the
curvature coordinates:
\begin{equation}
\bar{T}=\ln|V| - \ln |U| \,,\;\;\; \bar{R}={\cal R}(UV)\,.
\label{eq:Kruskal-curv}
\end{equation}
We see that the inversion $U \mapsto -U\,,\,V \mapsto -V$ leaves
$\bar{T}$ and $\bar{R}$ unchanged. Two points in the Kruskal plane
are labeled by the same curvature coordinates.

     It is useful to describe the mapping (\ref{eq:Kruskal-curv}) in
a slightly weaker form. Introduce the {\em tortoise coordinate}
\begin{equation}
\bar{R}^{\ast} = \bar{R} + \ln |\,1 - \bar{R} \,| \label{eq:turtleR}
\end{equation}
and combine $\bar{T}$ and $\bar{R}^{\ast}$ into null coordinates
\begin{equation}
\bar{U} = \bar{T} - \bar{R}^{\ast}, \;\;\; \bar{V} = \bar{T} +
\bar{R}^{\ast}.
\end{equation}
By multiplying and dividing the two equations (\ref{eq:UV}) we
obtain
\begin{equation}
U^{2}= \exp \big( - \bar{U} \big)\,, \;\;\; V^{2} = \exp \big(
\bar{V} \big) \,.  \label{eq:last}
\end{equation}

     The Schwarzschild line element expressed in the Kruskal
coordinates is everywhere regular, except at the initial and final
curvature singularities at $UV = 1 $.

\subsection{Geometrodynamical Description}

Geometrodynamics views a given spacetime as a dynamical evolution of
a three-geometry. Let us apply this viewpoint to the Schwarzschild
solution (\ref{eq:Schw}). Take an arbitrary spherically symmetric
spacelike hypersurface and let the spacetime metric induce on it the
spatial geometry $g$. Different hypersurfaces carry different
induced metrics.  As we evolve the spacetime metric along a
foliation which covers the entire Kruskal diagram, we build the
Schwarzschild solution.

     The simplest evolution is obtained along a one-parameter family
of hypersurfaces $T=const$ cutting across the regions I and III of
the Kruskal diagram. The geometry induced on all of these
hypersurfaces is the same: it is a wormhole geometry known under the
name of the Einstein-Rosen bridge. As we change $T$ from $-\infty$
to $\infty$, the geometry does not change at all.  It would be more
appropriate to speak in this case about geometrostatics rather than
geometrodynamics.  This reflects the fact that $\partial / \partial
T$ is a Killing vector field, and the $T=const$ hypersurfaces are
Lie-propagated by it. This evolution has a serious defect
\cite{Wheeler2}: The spacelike hypersurfaces $T=const$ do not cover
the entire Kruskal diagram, but only the static regions I and III\@.
The progress of time is arrested at the bifurcation point through
which all the hypersurfaces pass. The hypersurfaces thus do not form
a foliation.  Moreover, as the hypersurfaces proceed from past to
future in region I, they recede from future to past in region III\@.
The evolution does not proceed everywhere from past to future.

     The dynamical regions II and IV which were not covered by the
spacelike hypersurfaces $T=const$ can be covered by another simple
family of spacelike hypersurfaces, namely, $R=const <2M$. Their
geometry is again a wormhole geometry, but this time it describes a
homogeneous cylinder S$^{2} \times {\rm I\!R}$. As $R$ progresses
from $0$ to $2M$ in region IV, the cylinder opens up from the line
singularity, its circumference grows larger and larger while its
length per unit $T$ shrinks, and finally degenerates into a disk of
circumference $4 \pi M$ as $R$ approaches the horizon $R=2M$. In
region II, as $R$ decreases from $2M$ to $0$, the whole process is
reversed. Spatial geometry is dynamical in regions IV and II\@.  The
described evolution is locally isomorphic to the dynamics of the
Kantowski-Sachs universe \cite{Kantowski}. Unfortunately, the
foliations $0<R=const\,<2M$ are asymptotically null at infinities,
they approach the horizon for $R \rightarrow 2M$, and again, they do
not cover the whole Kruskal diagram. What they miss are exactly the
static regions.

     These shortcomings set our task. We want to study the spatial
geometry on a hypersurface which is spacelike and cuts the Kruskal
diagram all the way through: it starts at left infinity, goes
through the static region III, crosses the horizon into a dynamical
region, traverses it until it again reaches the horizon, crosses the
horizon to the static region I, and continues to right infinity. At
infinities, such a hypersurface should be asymptotically spacelike,
approaching some static hypersurface $T=T_{-}=const$ at left
infinity, and in general some other static hypersurface,
$T=T_{+}=const$, at right infinity.  One can cover the whole Kruskal
diagram by a foliation of such hypersurfaces. Even better, one can
admit all of them at once, and work in {\em many-fingered time
formalism}.

\section{Canonical Formalism for Spherically Symmetric Spacetimes}

     The geometrodynamical approach does not start from the known
Schwarzschild solution: it generates the Schwarzschild solution by
evolving a spherically symmetric geometry. The evolution is governed
by the Dirac-ADM action. In this section, we introduce the Dirac-ADM
action, and carefully discuss the necessary boundary conditions.

\subsection{Hypersurface Lagrangian $L_{\Sigma}$}

Take a spherically symmetric three-dimensional Riemannian space
$(\Sigma, g)$ and adapt the coordinates $x^{a}$ of its points $x \in
\Sigma$ to the symmetry: $x^{a}=(r, \theta, \phi)$. The line element
$d \sigma$ on $\Sigma$ is completely characterized by two functions,
$\Lambda (r)$ and $R(r)$, of the radial label $r$:
\begin{equation}
d \sigma ^{2} = \Lambda ^{2}(r)\,d r^{2} + R^{2}(r)\,d \Omega
^{2}\,.  \label{eq:lin-el}
\end{equation}
Again, $d \Omega $ is the line element on the unit sphere.

     Primordial black holes have the topology $\Sigma = {\rm I\!R}
\times {\rm S}^{2}$, and $r \in {\rm I\!R}$ thus ranges from
$-\infty$ to $\infty$. The coefficients $\Lambda (r)$ and $R(r)$
cannot vanish because the line element must be regular. We take both
of them to be positive, $\Lambda(r) > 0$ and $R(r) > 0\,$.  Then,
$R(r)$ is the curvature radius of the 2-sphere $r = const \,$, and
$d \sigma =\Lambda (r) dr$ is the radial line element oriented from
the left infinity to right infinity.

     Under transformations of $r$, $R(r)$ behaves as a scalar, and
$\Lambda (r)$ as a scalar density. This will simplify the momentum
constraint. Keeping this goal in mind, we have avoided the usual
exponential form of the metric coefficients. It is important to keep
track of the density character of the fundamental canonical
variables.  We shall always denote those canonical coordinates which
are scalars by Latin letters, and those which are scalar densities
by Greek letters.

     The line element (\ref{eq:lin-el}) leads to the curvature
scalar
\begin{equation}
{\bf R} [ g ] = -4 \Lambda ^{-2} R^{-1} R'' + 4 \Lambda ^{-3} R^{-1}
\Lambda ' R' - 2 \Lambda ^{-2} R^{-2} R'^{2} + 2 R^{-2} \, .
\label{eq:R}
\end{equation}

     Let us foliate a spherically symmetric spacetime $\cal M$ by
spherically symmetric leaves $\Sigma$, and label the leaves by a
time parameter $t \in {\rm I\! R}$. The metric coefficients
$\Lambda$ and $R$ then depend not only on $r$, but also on $t$. The
leaves of the foliation are related by the familiar lapse function
$N$ and the shift vector $N^{a}$.  Due to the spherical symmetry,
only the radial component $N^{r}$ of the shift vector survives, and
both $N(t,r)$ and $N^{r}(t,r)$ depend solely on the $t,r$ variables.

     The extrinsic curvature $K_{ab}$ of the leaves is given by the
rate of change $\dot {g} _{ab}$ of the metric with the label time $t
\,$:
\begin{equation}
K_{ab} = \frac{1}{2N} \big( - \dot {g} _{ab} + N_{(a|b)} \big) \, .
\end{equation}
For the spherically symmetric line element (\ref{eq:lin-el}),
$K_{ab}$ is diagonal, with
\begin{eqnarray}
K_{rr} & = &- N^{-1} \Lambda \Big( \dot{\Lambda} - \big( \Lambda
N^{r} \big)' \Big) \,, \label{eq:K-rr} \\ K_{\theta \theta} & = & -
N^{-1} R \Big( \dot{R} - R' N^{r} \Big) \,,\;\;\; K_{\phi \phi} =
\sin ^{2} \theta \, K_{\theta \theta}\, .  \label{eq:K-th}
\end{eqnarray}

     The vacuum dynamics of the metric field follows from the ADM
action
\begin{equation}
S_{\Sigma}[g,N,N^{a}] = \int dt \int _{\Sigma} d ^{3}x\, L_{\Sigma}
\label{eq:S}
\end{equation}
whose Lagrangian $L_{\Sigma}$ is
\begin{equation}
L_{\Sigma} = (16 \pi)^{-1} \, N |g|^{1/2} ( K^{ab}K_{ab} - K^{2} +
{\bf R}[g] ) \, . \label{eq:L}
\end{equation}
(In natural units $G=1=c$, the Einstein constant $\kappa = 8 \pi G /
c^{4}$ reduces to $ 8 \pi$.)  In a spherically symmetric spacetime
(\ref{eq:lin-el}), (\ref{eq:K-rr}), (\ref{eq:K-th})
\begin{eqnarray}
 \lefteqn{N|g|^{1/2}\,(K^{ab}K_{ab} - K^{2}) = -2N^{-1} \sin \theta}
\nonumber \\ & & \times \left( 2 \big( - \dot{\Lambda} + (\Lambda
N^{r})'\big) \big( - \dot{R} + R' N^{r} \big) R + \big( - \dot{R} +
R' N^{r} \big) ^{2} \Lambda \right) \,.
\end{eqnarray}
Integration over $\theta$ and $\phi$ gives the ADM action of a
Schwarzschild black hole:
\begin{eqnarray}
\lefteqn{S_{\Sigma}[R, \Lambda ;\, N, N^{r}] = \int dt \int _{-
\infty} ^{\infty} dr \,} \nonumber \\ & & \Bigg[ -\, N^{-1} \left( R
\big( - \dot{\Lambda} + (\Lambda N^{r})' \big) \big( - \dot{R} + R'
N^{r} \big) + \frac{1}{2} \Lambda \big( - \dot{R} + R' N^{r} \big)
^{2} \right) \nonumber \\ & & + \, N \left( - \Lambda ^{-1} R R'' +
\Lambda ^{-2} R R' \Lambda' - \frac{1}{2} \Lambda ^{-1} R'^{2} +
\frac{1}{2} \Lambda \right) \Bigg] \, .  \label{eq:S-lag}
\end{eqnarray}

     We shall discuss the appropriate boundary terms after passing
to the Hamiltonian formalism.

\subsection{Canonical Form of the Action}

By differentiating the ADM action with respect to the velocities
$\dot{\Lambda}$ and $\dot{R}$ we obtain the momenta
\begin{eqnarray}
P_{\Lambda} & = & -N^{-1} \, R \left( \dot{R} - R' N^{r} \right) \,,
\label{eq:PLambda} \\ \Pi _{R} & = & -N^{-1} \left( \Lambda \big(
\dot{R} - R' N^{r} \big) + R \big( \dot{\Lambda} - (\Lambda N^{r})'
\big) \right) \, .  \label{eq:PiR}
\end{eqnarray}
Throughout this paper, we denote those canonical coordinates which
are spatial scalars by Latin letters, and those which are spatial
densities by Greek letters. The conjugate momenta always carry
complementary weights. Therefore, the momentum $P_{R}$ conjugate to
the scalar $R$ is a density, while the momentum $P_{\Lambda}$
conjugate to the density $\Lambda$ is a scalar.

     Equations (\ref{eq:PLambda}) and (\ref{eq:PiR}) can be inverted
for the velocities,
\begin{eqnarray}
\dot{\Lambda} & = & -N R^{-2} \big( R \Pi_{R} - \Lambda P_{\Lambda}
\big) + \big( \Lambda N^{r} \big)' \,, \\ \dot{R} & = &
-N\,R^{-1}P_{\Lambda} + R' N^{r} \,. \label{eq:dotR}
\end{eqnarray}
They allow us to write the extrinsic curvature as a function of the
canonical momenta:
\begin{equation}
K_{rr} = \Lambda R^{-2} \big( R \Pi_{R} - \Lambda P_{\Lambda} \big)
\,, \;\; K_{\theta \theta} = P_{\Lambda} \, . \label{eq:K-P}
\end{equation}

By symmetry, the curvature of a normal section of $\Sigma$ attains
its extremal values $K_{1}, K_{2}, K_{3}$ (called {\em principal
curvatures}) for those sections which are either tangential
$\parallel$ or normal $\perp$ to the 2-spheres $r=const$. The last
equation enables us to express these principal curvatures
$K_{1}=K_{\parallel}$ and $K_{2}=K_{3}=K_{\perp}$ in terms of the
momenta:
\begin{eqnarray}
K_{\parallel} & = & K^{r}_{r} = \Lambda^{-1} R^{-2} \big( R \Pi_{R}
- \Lambda P_{\Lambda} \big) \,, \nonumber \\ K_{\perp} & = &
K^{\theta}_{\theta} = K^{\phi}_{\phi} = R^{-2} P_{\Lambda}\,.
\end{eqnarray}
Inversely,
\begin{equation}
P_{\Lambda} = R^{2} K_{\perp}\,,\;\; \Pi_{R} = R
\Lambda\,\big(K_{\parallel}+ K_{\perp} \big)\,.
\end{equation}
This endows the canonical momenta with an invariant geometric
meaning.  Note that $P_{\Lambda}$ is proportional to $K_{\perp}$,
but $\Pi_{R}$ is proportional to the {\em sum} of $K_{\parallel}$
and $K_{\perp}\,$, not to $K_{\parallel}$ itself. This sum also
differs from the {\em mean curvature}
\begin{equation}
K\; \mbox{$:=$}\; K_{1}+K_{2}+K_{3}= K^{a}_{a} = K_{\parallel} +
2K_{\perp}\, .
\end{equation}

     The action (\ref{eq:S-lag}) can be cast into the canonical form
by the Legendre dual transformation:
\begin{equation}
S_{\Sigma}[\Lambda, R, P_{\Lambda},\Pi_{R};\, N, N^{r}] = \int dt
\int _{-\infty}^{\infty}dr \, \big( P_{\Lambda} \dot{\Lambda} +
\Pi_{R} \dot{R} - NH -N^{r}H_{r} \big) \, . \label{eq:S-ham}
\end{equation}
By this process, we obtain the super-Hamiltonian
\begin{eqnarray}
\lefteqn {H \, \mbox{$:=$} \, -R^{-1} \Pi_{R} P_{\Lambda} +
\frac{1}{2} R^{-2} \Lambda P_{\Lambda}^{2}} \nonumber \\ & & \mbox{}
+ \Lambda^{-1} R R'' - \Lambda^{-2} R R' \Lambda' + \frac{1}{2}
\Lambda ^{-1} R'^{2} - \frac{1}{2} \Lambda \label{eq:SH}
\end{eqnarray}
and supermomentum
\begin{equation}
H_{r} \, \mbox{$:=$}\, \Pi_{R} R' - \Lambda P_{\Lambda}' \, .
\label{eq:SM}
\end{equation}

The form of $H_{r}$ is dictated by the requirement that it generate
Diff$\,{\rm I\!R}$ of the scalars $R$ and $P_{\Lambda}$, and of the
scalar densities $\Pi_{R}$ and $\Lambda$. The minus sign in
({\ref{eq:SM}) is due to the fact that the momentum $P_{\Lambda}$ is
a scalar, and the coordinate $\Lambda$ a scalar density, rather than
the other way around.

     The expressions (\ref{eq:SH}) and (\ref{eq:SM}) can be obtained
from those derived by BCMN \cite{Berger} by a point transformation.

\subsection{Falloff of the Canonical Variables}

So far, we paid no attention to the behavior of the canonical
variables $\Lambda, R$ and $ P_{\Lambda}, \Pi_{R}$ at infinity. The
importance of the falloff conditions was pointed out by Regge and
Teitelboim \cite{RT}, \cite{HRT}, and their form refined by many
authors.  We shall follow the treatment of Beig and O'Murchadha
\cite{BO'M}.

     Primordial black holes have two spatial infinities rather than
just one. We shall formulate the falloff conditions at right
infinity, and then state what the corresponding conditions are at
left infinity.

     Let $x^{a}$ be a global system of coordinates on $\Sigma$ which
is asymptotically Cartesian. Such a system is related to the
spherical system of coordinates $r, \theta, \phi$ by the standard
flat space formulae. At $r \rightarrow \infty$, the metric $g_{ab}$
and the conjugate momentum $p^{ab}$ are required to fall off as
\begin{eqnarray}
g_{ab}(x^{c}) & = & \delta_{ab} + r^{-1} h_{ab}(n^{c}) + O^{\infty}
(r^{-(1+\epsilon)})\,, \label{eq:gfall} \\ p^{ab}(x^{c}) & = &
r^{-2} k^{ab}(n^{c}) + O^{\infty} (r^{-(2+\epsilon)}) \,.
\label{eq:pifall}
\end{eqnarray}
Here, $n^{a}\, \mbox{$:=$}\, x^{a}/r$, and
$f(x^{a})=O^{\infty}(r^{-n})$ means that $f$ falls off like
$r^{-n}$, $f_{,a}$ like $r^{-(n+1)}$, and so on for all higher
spatial derivatives. The leading term in $g_{ab} - \delta_{ab}$ is
of the order $r^{-1}$, and the leading term in $p^{ab}$ is one order
higher, $r^{-2}$. The coefficients $h_{ab}$ and $k^{ab}$ are
required to be smooth functions on ${\rm S}^{2}$, and to have
opposite parities: $h_{ab}(n^{c})$ is to be even, and
$k^{ab}(n^{c})$ is to be odd, i.e.\,
\begin{equation}
h_{ab}(-n^{c}) = h_{ab}(n^{c})\,, \;\; k^{ab}(-n^{c}) =
-k^{ab}(n^{c})\,.
\end{equation}
Together with the canonical data, the lapse function and the shift
vector are assumed to behave as
\begin{eqnarray}
N(x^{c}) & = & N_{+}(n^{c}) + O^{\infty} (r^{-\epsilon}) \, , \\
N^{a}(x^{c}) & = & N_{+}^{a}(n^{c}) + O^{\infty} (r^{-\epsilon})
\end{eqnarray}
at infinity.

     Because $dr = n_{a}dx^{a}$ and $d \Omega^{2} = r^{-2}
(\delta_{ab} - n_{a}n_{b})\,dx^{a}dx^{b}$, the spherically symmetric
metric (\ref{eq:lin-el}) can easily be transformed into Cartesian
coordinates:
\begin{eqnarray}
g_{ab} & = & \Lambda^{2} n_{a}n_{b} + \left( \frac{R}{r} \right)
^{2} \,\big( \delta_{ab} - n_{a}n_{b} \big) \, , \label{eqn:gcart}
\\ g^{ab} & = & \Lambda^{-2} n^{a}n^{b} + \left( \frac{r}{R}
\right)^{2} \big( \delta^{ab} - n^{a}n^{b} \big) \, .
\label{eq:gcart'}
\end{eqnarray}
The general falloff conditions (\ref{eq:gfall}) then determine the
behavior of the metric coefficients $R$ and $\Lambda$ at infinity:
\begin{equation}
\Lambda(t,r) = 1 + M_{+}(t)r^{-1} + O^{\infty}(r^{-(1+\epsilon)})\,
,
\end{equation}
and
\begin{equation}
R(t,r) = r + \rho_{+}(t) + O^{\infty}(r^{-\epsilon}) \, .
\end{equation}
Here, due to spherical symmetry, $M_{+}(t)$ and $\rho_{+}(t)$ cannot
depend on the angles $n^{c}$, but they can still depend on $t$. Of
course, $M_{+}$ is the Schwarzschild mass as observed at right
infinity. The general falloff conditions allow $R$ to differ by the
amount $\rho_{+}(t)$ from $r$ at infinity . Because we want $r$ to
coincide with $R$ at infinity, we impose stronger falloff
conditions on $R$ by putting $\rho_{+}=0\,$:
\begin{equation}
R(r) = r + O^{\infty}(r^{-\epsilon})\, .
\end{equation}
Consistency then demands that the shift vector must also
asymptotically vanish, $N^{r}_{+} = 0\,$:
\begin{equation}
N^{r}(r) = O^{\infty}(r^{-\epsilon}) \, .
\end{equation}
Unlike $N^{r}$, the lapse function cannot vanish at infinity. If it
did, the time there would stand still. Instead, $N(t,r)$ assumes at
infinity an angle-independent value $N_{+}(t)\,$:
\begin{equation}
N(r) = N_{+}(t) + O^{\infty}(r^{-\epsilon})\, .
\end{equation}

     Our next task is to determine the falloff of the canonical
momenta.  The second fundamental form $K_{ab}dx^{a}dx^{b}$ can be
transformed into Cartesian coordinates by the same procedure as the
metric. By using (\ref{eq:K-P}) we get
\begin{eqnarray}
\lefteqn {K_{rr} dr^{2} + K_{\theta \theta} d\theta^{2} + K_{\phi
\phi} d\phi^{2} =} \nonumber \\ & & \Lambda R^{-2} \big( R \Pi_{R} -
\Lambda P_{\Lambda} \big)\, dr^{2} + P_{\Lambda}\, d\Omega^{2} =
\nonumber \\ & & \Big( \Lambda R^{-2} \big( R \Pi_{R} - \Lambda
P_{\Lambda} \big) n_{a} n_{b} + r^{-2} P_{\Lambda} \big( \delta_{ab}
- n_{a}n_{b} \big) \Big) \, dx^{a}dx^{b}\,.
\end{eqnarray}
{}From here we read the extrinsic curvature and, by referring back to
the metric (\ref{eq:gcart'}), find the canonical momentum
\begin{eqnarray}
\lefteqn{p^{ab}\,\mbox{$:=$}\,|g|^{1/2} \big( Kg^{ab} - K^{ab} \big)
} \nonumber \\ & & = 2r^{-2} \Lambda^{-1} P_{\Lambda} n^{a}n^{b} +
R^{-1} \Pi_{R} \big( \delta^{ab} - n^{a}n^{b} \big) \, .
\end{eqnarray}
This expression for $p^{ab}$ is even in the angular variables.  This
means that $k^{ab}$ is even. The requirement that $k^{ab}$ be odd
implies that $k^{ab}$ must vanish, i.e.\
\begin{equation}
P_{\Lambda} = O^{\infty}(r^{-\epsilon})\, , \;\; \Pi_{R} =
O^{\infty}(r^{-(1+\epsilon)}) \, .
\end{equation}

     The same considerations apply at $r \rightarrow -\infty \,$.
Taken together, they imply the falloff conditions
\begin{eqnarray}
\Lambda (t,r) &=& 1 + M_{\pm}(t) |r|^{-1} +
O^{\infty}(|r|^{-(1+\epsilon)})\,, \label{eq:Lambdafall} \\ R(t,r)
&=& |r| + O^{\infty}(|r|^{-\epsilon}) \, , \label{eq:Rfall} \\
P_{\Lambda}(t,r) & = & O^{\infty}(|r|^{-\epsilon}) \, ,
\label{eq:PiLambdafall} \\ \Pi_{R}(t,r) & = &
O^{\infty}(|r|^{-(1+\epsilon)}) \label{eq:PiRfall}
\end{eqnarray}
for the canonical variables, and the falloff conditions
\begin{eqnarray}
N(t,r) & = & N_{\pm}(t) +
O^{\infty}(|r|^{-\epsilon})\,,\label{eq:Nfall} \\ N^{r}(t,r) & = &
O^{\infty}(|r|^{-\epsilon}) \label{eq:Nrfall}
\end{eqnarray}
for the Lagrange multipliers.

     Let us note again that $R(r)$ approaches $|r|$ at the rate
$O^{\infty}(|r|^{-\epsilon})$ as $r \rightarrow \pm \infty$.  After
we reconstruct the Killing time $T(r)$ from the canonical data, we
shall prove a similar result for the approach of the $t = const$
foliation to the $T = const$ foliation:
\begin{equation}
T(t,r) = T_{\pm}(t) + O^{\infty}(|r|^{-\epsilon}) \, .
\label{eq:Tfall}
\end{equation}

     The falloffs of the canonical data ensure that the Liouville
form
\begin{equation}
\int ^{\infty}_{-\infty} dr \, \big( P_{\Lambda} \dot{\Lambda} +
\Pi_{R} \dot{R}\, \big)
\end{equation}
is well defined. They also imply that the super-Hamiltonian and
supermomentum fall off as
\begin{equation}
H = O^{\infty}(|r|^{-(1+\epsilon)}) \,, \;\; H_{r} =
O^{\infty}(|r|^{-(1+\epsilon)}) \, . \label{eq:Hfall}
\end{equation}
Equations (\ref{eq:Nfall}), (\ref{eq:Nrfall}) and (\ref{eq:Hfall})
ensure that the Hamiltonian is well defined. The canonical action
$S_{\Sigma}$ thus has a good meaning.

     The falloff conditions for spherically symmetric {\em vacuum}
spacetimes may easily be strengthened. The only necessary condition
on how fast the canonical variables fall down is that the
Schwarzschild solution itself can be made to fall down that fast by
an appropriate choice of the hypersurface and its parametrization
$r$ at infinities.  By choosing the hypersurface to coincide with a
$T=t$ hypersurface outside a spherical tube $R = R_{0} = const\,$,
and by labeling it there by the curvature coordinate, $R = |r|\,$,
we achieve that
\begin{equation}
\begin{array}{ll}
\Lambda (t,r) = 1 + M_{\pm}(t)|r|^{-1}\,,& R(t,r) = |r|\,, \\
P_{\Lambda}(t,r) = 0\,, & \Pi_{R}(t,r) = 0 \label{eq:strongfall}
\end{array}
\end{equation}
for the Schwarzschild solution. We are thus free to require that our
phase space variables satisfy (\ref{eq:strongfall}) everywhere
outside a compact region.

\subsection{Boundary Terms}

Arnowitt, Deser and Misner complemented the hypersurface action $
S_{\Sigma}$ by a boundary action $S_{\partial \Sigma}$ at infinity:
\begin{equation}
S = S_{\Sigma} + S_{\partial \Sigma} \,. \label{eq:Stot}
\end{equation}
A primordial black hole has two infinities, and hence there are two
boundary contributions
\begin{equation}
S_{\partial \Sigma} = - \int dt \, \big( N_{+}(t)E_{+}(t) +
N_{-}(t)E_{-}(t) \big) \, . \label{eq:Sbound}
\end{equation}
Each of them is the product of the lapse function with the ADM
energy
\begin{equation}
E_{\pm} = (16 \pi)^{-1} \int_{{\rm S}^{2}_{\pm}} dS^{a} \,
\delta^{bc} \big( g_{ab,c} - g_{bc,a} \big)
\end{equation}
which is given by an integral over a two-sphere ${\rm S}^{2}_{\pm}$
at infinity. In the asymptotically Cartesian coordinates
\begin{equation}
dS^{a} = r^{2}n^{a}\sin \theta \, d \theta d \phi \,, \;\; |r|_{,a}
= n_{a}\,, \;\; n_{a,b} = |r|^{-1}(\delta_{ab} - n_{a}n_{b}) \, ,
\end{equation}
and the asymptotic form (\ref{eq:gfall}) of the metric yields
\begin{equation}
E_{\pm} = \lim_{r \rightarrow \pm \infty} \frac{1}{2} |r| \, \big(
\Lambda^{2} - 1 \big) = M_{\pm}(t) \, .  \label{eq:ADMenergy}
\end{equation}
The ADM energy of a black hole is its Schwarzschild mass.

     The total action of a primordial black hole takes the form
\begin{eqnarray}
\lefteqn {S[\Lambda, P_{\Lambda}, R, \Pi_{R};\, N, N^{r}] = }
\nonumber \\ & & \int dt \int _{-\infty}^{\infty} dr \, \big(
P_{\Lambda} \dot{\Lambda} + \Pi_{R} \dot{R} - NH - N^{r}H_{r} \big)
\nonumber \\ & & - \int dt\, (N_{+}M_{+} + N_{-}M_{-}) \, .
\label{eq:Sfin}
\end{eqnarray}

     Variations of the canonical variables should preserve the
prescribed falloffs. In particular, the leading term in the
variation of $\Lambda$ is given by the variation of the
Schwarzschild mass:
\begin{equation}
\delta \Lambda _{\pm} = \delta M_{\pm} \, |r|^{-1} + O^{\infty}
(|r|^{-2}) \, .
\end{equation}
As emphasized by Regge and Teitelboim \cite{RT}, without the ADM
boundary action this particular variation of the action would lead
to an inconsistency. On the other hand, when the ADM boundary action
is included, another inconsistency would arise if one allowed the
variation of the lapse function at infinities. It is therefore
important to treat $N_{\pm}(t)$ as {\em prescribed functions} of
$t$.

     To see why this is so, let us identify those terms in the
Hamiltonian density $NH + N^{r}H_{r}$ whose variations lead to
boundary terms. The variation of $\Pi_{R}$ does not yield any
boundary term because there are no derivatives of $\Pi_{R}$ anywhere
in the action. The variables $P_{\Lambda}$ and $R$ also do not cause
any trouble, because the falloff conditions ensure that the
boundary terms brought in by the variation of $P_{\Lambda}$ and $R$
safely vanish.  The sole troublemaker is the derivative term $NRR'
\Lambda^{-2} \Lambda '$ in $-NH$. Its variation with respect to
$\Lambda$ yields the boundary term
\begin{equation}
N_{+}(t) \delta M_{+}(t) + N_{-}(t) \delta M_{-}(t)\, .
\label{eq:bound}
\end{equation}
If there were no ADM boundary action, the variation of the
hypersurface action $S_{\Sigma}$ with respect to $\Lambda$ would
lead to the conclusion that $N_{\pm}(t) = 0 \,$, i.e.\, it would
freeze the evolution at infinities.  The ADM boundary action is
designed so that its variation with respect to $M_{\pm}(t)$ exactly
cancels the boundary term (\ref{eq:bound}), and the unwanted
conclusion does not follow.

     The ADM boundary action leads, however, to an inconvenient
caveat.  Without it, the lapse and the shift functions in the
hypersurface action $S_{\Sigma}$ can be freely varied at infinities
because the super-Hamiltonian $H$ and the supermomentum $H_{r}$
asymptotically vanish. When the boundary action is included, the
variation of the total action $S = S_{\Sigma} + S_{\partial \Sigma}$
yields
\begin{equation}
\delta_{N} S = - \int dt\, \big( M_{+}(t) \delta N_{+}(t) + M_{-}(t)
\delta N_{-} \big) \,.
\end{equation}
If we allowed $N(t,r)$ to be varied at infinities, we would get an
unwanted `natural boundary condition' $M_{+} = 0 = M_{-}$. This
would exclude black hole solutions and leave only a flat spacetime.
Therefore, in the variational principle (\ref{eq:Sfin}) we must
demand that the values $N_{\pm}(t)$ of the lapse function at
infinities be some {\em prescribed} functions of $t$ which cannot be
varied. This means that the lapse function $N$ in our variational
principle has {\em fixed ends}.

\subsection{Parametrization at Infinities}

     The necessity of fixing the lapse func\-tion at infini\-ties
can be removed by {\em paramet\-rization}. The lapse function is the
rate of change of the proper time $\tau$ with respect to the label
time $t$ in the direction normal to the foliation. Because
$N^{r}_{\pm}(t) = 0$, we can write
\begin{equation}
N_{\pm}(t) = \pm \dot{\tau} _{\pm}(t) \, , \label{eq:Npar}
\end{equation}
where $\tau _{\pm}(t)$ is the proper time measured on standard
clocks moving along the $r=const$ worldlines at infinities. By
convention, we let the proper time at the left infinity decrease
from the past to the future, to match the behavior of the Killing
time $T$ in the Kruskal diagram. This introduces the minus sign in
(\ref{eq:Npar}) at $-\infty$.

     We now replace the lapse function in the ADM boundary action by
the derivatives of $\tau_{\pm}$, (\ref{eq:Npar}), and treat
$\tau_{\pm}(t)$ as additional variables:
\begin{eqnarray}
\lefteqn{ S[\Lambda, P_{\Lambda}, R, \Pi_{R};\, N, N^{r};\,
\tau_{+}, \tau_{-}] =} \nonumber \\ & & \int dt
\int_{-\infty}^{\infty} dr\, \big( P_{\Lambda} \dot{\Lambda} +
\Pi_{R} \dot{R} - NH - N^{r}H_{r} \big) \nonumber \\ & & - \int dt
\, \big( M_{+} \dot{\tau}_{+} - M_{-} \dot{\tau}_{-} \big) \,.
\label{eq:Spa}
\end{eqnarray}
This rearrangement of the action is called {\em the parametrization
at infinities}. Notice that $N_{\pm}$ still appears in the
hypersurface part $S_{\Sigma}$ of the action.

     The variables $\tau_{\pm}(t)$ in the action (\ref{eq:Spa}) can
be freely varied. The result of their variation is a valid equation,
namely, the mass conservation
\begin{equation}
\dot{M}_{+}(t) = 0 = \dot{M}_{-}(t) \,.
\end{equation}
The lapse function at infinities $N_{\pm}(t)$ can also be freely
varied, because it now occurs only in the hypersurface action, and
the super-Hamiltonian asymptotically vanishes.

     The only remaining question is what happens under the variation
of $ \Lambda$. As before, the variation of the hypersurface action
$S_{\Sigma}$ gives the boundary term (\ref{eq:bound}). On the other
hand, the variation of the parametrized ADM boundary action in
(\ref{eq:Spa}) now yields
\begin{equation}
- \dot{\tau}_{+}(t) \delta M_{+}(t) + \dot{\tau}_{-}(t) \delta
M_{-}(t)\,.
\end{equation}
Before parametrization, the variation of $\Lambda$ at infinities,
i.e.\ the variation of $M_{\pm}(t)$, produced merely an identity.
After parametrization, the situation is different: the variation of
$M_{\pm}(t)$ relates $\tau_{\pm}(t)$ to $N_{\pm}(t)$ by
(\ref{eq:Npar}). These equations thereby follow as {\em natural
boundary conditions} from the parametrized action principle
(\ref{eq:Spa}).

     As we proceed, we shall at first pay little attention to the
ADM boundary action. We shall meet it again in
Section~\ref{sec-tale}.

\section{Reconstructing the Mass and Time from the Canonical Data}
\label{sec-MT}

\subsection{Reconstruction Program}

In canonical gravity, we know the intrinsic metric and extrinsic
curvature of a hypersurface, but we do not have any {\em a priori}
knowledge about how the hypersurface is located in spacetime.
Suppose that we are given the canonical data $\Lambda, R,
P_{\Lambda}, \Pi_{R}$ on a spherically symmetric hypersurface
cutting across a Schwarzschild black hole. Can we tell from those
data the mass of the black hole?  And can we infer from them how the
hypersurface is drawn in the Kruskal diagram?

     Disregarding some subtleties concerning the anchoring of the
hypersurface at infinities, the answer to both of these questions is
{\em yes}. To arrive at the answers, we start from the knowledge
that the hypersurface must ultimately be embedded in a spacetime
endowed with the line element (\ref{eq:Schw}). Let the hypersurface
be a leaf of a foliation
\begin{equation}
T = T(t,r)\,, \;\; R = R(t,r)\,. \label{eq:fol}
\end{equation}
The line element (\ref{eq:Schw}) induces an intrinsic metric and
extrinsic curvature on the hypersurface. By comparing these
quantities with the canonical data, we connect the spacetime
formalism with the canonical formalism. This connection enables us
to identify the Schwarzschild mass $M$ and the embedding
(\ref{eq:fol}) from the canonical data.

     We can expect discontinuities at the horizon where different
patches of the curvature coordinates $T$ and $R$ meet each other.
Indeed, $T$ becomes infinite at the horizon. We shall see, however,
that the transition of $T$ across the horizon is well under our
control, and can be predicted from smooth canonical data. We can
then easily pass from the curvature coordinates $T, R$ to the
Kruskal coordinates $U, V$ which are continuous across the horizon.
The direct reconstruction of the Kruskal coordinates would be much
more cumbersome.

     To implement our program, we substitute the foliation
(\ref{eq:fol}) into the Schwarz\-schild line element (\ref{eq:Schw})
and get
\begin{eqnarray}
\lefteqn{ds^{2} = - \big( F \dot{T}^{2} - F^{-1} \dot{R}^{2} \big)\,
dt^{2}} \nonumber \\ & & \mbox{} + 2\big( - F T' \dot{T} + F^{-1} R'
\dot{R} \big)\, dt dr \nonumber \\ & & \mbox{} + \big( - F T'^{2} +
F^{-1} R'^{2} \big)\, dr^{2} + R^{2} d\Omega^{2}.
\label{eq:schw-tr}
\end{eqnarray}
By comparing this result with the ADM form of the line element
\begin{eqnarray}
\lefteqn{ ds^{2} = - \big( N^{2} - \Lambda^{2} (N^{r})^{2} \big)\,
dt^{2}} \nonumber \\ & & \mbox{} + 2 \Lambda^{2} N^{r} dt dr
\nonumber \\ & & \mbox{} + \Lambda^{2} dr^{2} + R^{2} d\Omega^{2}
\,, \label{eq:ADM-le}
\end{eqnarray}
we obtain a set of three equations,
\begin{eqnarray}
\Lambda^{2} &=& -F T'^{2} + F^{-1} R'^{2} \,, \label{eq:lambda} \\
\Lambda^{2} N^{r} &=& -F T' \dot{T} + F^{-1} R' \dot{R} \,,
\label{eq:forNr} \\ N^{2} - \Lambda^{2} (N^{r})^{2} &=& F\dot{T}^{2}
- F^{-1}\dot{R}^{2}.  \label{eq:forN}
\end{eqnarray}
The first two equations can be solved for $N^{r}$,
\begin{equation}
N^{r} = \frac{-F T' \dot{T} + F^{-1} R' \dot{R}} {-F T'^{2} + F^{-1}
R'^{2}} \, .  \label{eq:Nr}
\end{equation}
This solution $N^{r}$, together with $\Lambda$ of equation
(\ref{eq:lambda}), can be substituted into the remaining equation
(\ref{eq:forN}), which can then be solved for the lapse:
\begin{equation}
N = \frac{R' \dot{T} - T' \dot{R}} {\sqrt{-F T'^{2} + F^{-1}
R'^{2}}} \,.  \label{eq:N}
\end{equation}

     The lapse function $N$ should be positive. The transition from
(\ref{eq:forN}) to (\ref{eq:N}) requires taking a square root. We
must check that the square root we have taken is positive.

     First of all, the denominator of (\ref{eq:N}) is real and
positive: it is equal to $\Lambda$. We shall check that the
numerator of (\ref{eq:N}) is positive separately in each region of
the Kruskal diagram. Because the lapse function is a spatial scalar,
we first choose in each region an appropriate radial label $r\,$:
$r=R$ in region I, $r=T$ in region II, $r=-R$ in region III, and
$r=-T$ in region IV\@.  Under these choices, the numerator of
(\ref{eq:N}) becomes $\dot{T}$ in region I, $-\dot{R}$ in region II,
$-\dot{T}$ in region III, and $\dot{R}$ in region IV\@. With the
label time going to the future, all these expressions are positive.

    We now substitute the expressions (\ref{eq:Nr}) and (\ref{eq:N})
for $N^{r}$ and $N$ into (\ref{eq:PLambda}) and calculate
$P_{\Lambda}\,$.  The time derivatives $\dot{T}$ and $\dot{R}$
dutifully drop out, and we get the relation
\begin{equation}
-T' = R^{-1}F^{-1} \Lambda P_{\Lambda} \,.  \label{eq:T'}
\end{equation}
When we substitute this $T'$ back into (\ref{eq:lambda}), we can
calculate $F$ as a function of the canonical data:
\begin{equation}
F = \left( \frac{R'}{\Lambda} \right) ^{2} - \left(
\frac{P_{\Lambda}}{R} \right) ^{2} . \label{eq:canF}
\end{equation}
Taken together, the last two equations express $T'$ in terms of the
canonical data. Moreover, because we know $F$ in terms of $M$ and
$R$, (\ref{eq:F}), we can also determine the Schwarzschild mass
\begin{equation}
M = \frac{1}{2} R^{-1} P_{\Lambda}^{2} - \frac{1}{2} \Lambda^{-2} R
R'^{2} + \frac{1}{2} R \, . \label{eq:canM}
\end{equation}

     Equations (\ref{eq:T'}) - (\ref{eq:canM}) accomplish our goal.
Equation (\ref{eq:canM}) enables us to read the mass of the
Schwarzschild black hole from the canonical data on any small piece
of a spacelike hypersurface. It does not matter whether that piece
is close to or far away from infinities, or even whether it lies
inside or outside the horizon. Equations (\ref{eq:T'}) and
(\ref{eq:canF}) determine the {\em difference} of the Killing times
$T(r_{1})$ and $T(r_{2})$ between any two points, $r_{1}$ and
$r_{2}$, of the hypersurface. To determine $T(r)$ itself, we need to
know $T$ at one point of the hypersurface, say, at the right
infinity. The equations (\ref{eq:T'}) - (\ref{eq:canM}) are the key
to our treatment of the Schwarzschild black holes. We shall now
explore their consequences.

\subsection{Across the Horizon}

Our assertion that equations (\ref{eq:T'}) and (\ref{eq:canF})
determine the difference of the Killing times between any two points
of the hypersurface requires a caveat.  On the horizon, the
coefficient $F(r)$ vanishes, and the time gradient (\ref{eq:T'})
becomes infinite.  We must show that we can propagate our knowledge
of time across the horizon.

     A spacelike hypersurface must intersect both the leftgoing and
the rightgoing branches of the horizon. Unless it passes straight
through the bifurcation point, it has two intersections with the
horizon.  Inside the horizon, the hypersurface lies either entirely
within the future dynamical region, or entirely within the past
dynamical region.

     How does one recognize from the smooth canonical data where the
hypersurfaces crosses the horizon? One looks at the points where the
combination (\ref{eq:canF}) vanishes. The function (\ref{eq:canF})
can be written as a product of two factors,
\begin{equation}
F = F_{+} \times F_{-}\,, \;\;\; F_{\pm} \; \mbox{$:=$} \;
\frac{R'}{\Lambda} \pm \frac{P_{\Lambda}}{R}\,.
\end{equation}
At the horizon, at least one of the two factors must vanish.

     To find what branches of the horizon are described by the
factors $F_{\pm}$, we must determine the signs of several quantities
at the intersection of the hypersurface with the horizon. This is
easily done by inspecting the Kruskal diagram.  We chose our radial
label to grow from left infinity to right infinity.  Therefore,
inside the future dynamical region $T(r)$ increases with $r$,
$T'(r)>0$, and inside the past dynamical region it decreases with
$r$, $T'(r)<0$. As the hypersurface is entering a dynamical region
from the left static region, $R(r)$ is falling, $R'(r)<0$, and as it
is exiting the dynamical region into the right static region, $R(r)$
is growing, $R'(r)>0$. If the hypersurface passes through the
bifurcation point, both $R'(r)=0$ and $T'(r)=0$.

     In the dynamical regions, $F<0$, and equation (\ref{eq:T'})
tells us that $T'(r)$ and $P_{\Lambda}(r)$ have the same sign.
Therefore, $P_{\Lambda}(r)$ is positive in the future dynamical
region, and negative in the past dynamical region. From continuity,
$P_{\Lambda}(r)>0$ at intersections with the future horizon $\vee$,
negative at intersections with the past horizon $\wedge$, and zero
when the hypersurface crosses the bifurcation point.  We already
know that $R'(r)<0$ when the hypersurface dives from the left static
region through the $>$ part of the horizon into a dynamical region,
and $R'(r)>0$ when it reemerges through the $<$ part of the horizon
into the right static region. Putting these facts together, we see
that the equation $F_{+}(r)=0$ defines the leftgoing branch of the
horizon, and $F_{-}(r)=0$ defines the rightgoing branch. When both
$F_{+}(r)$ and $F_{-}(r)$ simultaneously vanish, the hypersurface
goes through the bifurcation point.

     We can now return to the problem of determining the passage of
time across the horizon. To be definite, let the hypersurface cross
the horizon from the future dynamical region into the right static
region at a point $r_{0}:\, R'_{0}>0,\, P_{\Lambda 0}>0$.  Because
$R'_{0} >0$, we can choose $R$ as a radial coordinate on the
hypersurface in the vicinity of the crossing. The crossing condition
$F_{-}(r_{0})=0$ then implies $(\Lambda P_{\Lambda})_{0}=R_{0}$, and
equation (\ref{eq:T'}) reduces to
\begin{equation}
\frac{dT(R)}{dR} = - \frac{1}{2} R_{0} \Lambda^{2}_{0}\,
\frac{1}{R-R_{0}}
\end{equation}
for $R$ close to $R_{0}$. From here we determine the transition of
$T(R)$ across the horizon:
\begin{equation}
T(R) = - \frac{1}{2} R_{0}\Lambda_{0}^{2}\, \ln |R-R_{0}| + const
\,.
\end{equation}
As expected, $T$ becomes infinite at the horizon, but in a well
determined rate. The value of the `constant' is determined by
matching $ T(R)$ at one side of the horizon to its given value.

     If the hypersurface passes through the bifurcation point,
$T(r)$ changes continuously and remains finite. We conclude that we
can determine the passage of $T(r)$ through the horizon from the
canonical data.

\section{Mass Function and Time Gradient as Canonical Variables}

     Equations (\ref{eq:T'}) - (\ref{eq:canM}) were obtained from
the known form of the Schwarzschild solution, i.e.\, by implicitly
using the Einstein equations. Let us now forget their humble origin,
and promote the expressions for $M(r)$ and $-T'(r)$ to {\em
definitions} of two sets of dynamical variables on our phase space.
Note that $M(r)$ is a local functional of the canonical data, and as
such it depends on $r$.  Indeed, prior to imposing the constraints
and the Hamilton equations on the data, $M(r)$ does not need to be
constant.

      We can interpret the function $M(r)$ as the mass content of
the wormhole to the left of the two-sphere labeled by $r$. The other
function, $-T'(r)$, tells us the rate at which the Killing time
$T(r)$ falls with $r$. A remarkable feature of these two functions
is that they form a pair of canonically conjugate variables.
Anticipating the outcome of the proof we are going to present, we
denote the dynamical variable $-T'(r)$ as $\Pi_{M}(r)$.  Our density
notation still applies: by its construction (\ref{eq:canM}), $M(r)$
is a spatial scalar, while
\begin{equation}
\Pi_{M} = R^{-1} F^{-1} \Lambda P_{\Lambda} \label{eq:PiM}
\end{equation}
is a scalar density.

     Because the expressions for $M(r)$ and $\Pi_{M}(r)$ do not
contain $\Pi_{R}$, they have vanishing Poisson brackets with $R(r)$.
Unfortunately, their Poisson brackets with $\Pi_{R}(r)$ do not
vanish.  We thus cannot complement the variables $M(r)$,
$\Pi_{M}(r)$ by the canonical pair $R(r)$, $\Pi_{R}(r)$, and get
thereby a new canonical chart on the phase space.

     Obviously, we need to modify the momentum $\Pi_{R}(r)$ in such
a way that the new momentum, $\Pi_{\sf R}(r)$, will commute with
$M(r)$ and $P_{M}(r)$, but still remains conjugate to
\begin{equation}
{\sf R} = R \,.  \label{eq:sfR}
\end{equation}
The only way of doing this is to add to $\Pi_{R}(r)$ a dynamical
variable $\Theta(r)$ that does not depend on $\Pi_{R}(r)$:
\begin{equation}
\Pi_{\sf R}(r) = \Pi_{R}(r) + \Theta (r; \, R, \Lambda,
P_{\Lambda}]\,.
\end{equation}
To guess the correct $\Theta$ is tricky. Our guiding principle is
that the variables $M(r), {\sf R};\, \Pi_{M}(r), \Pi_{\sf R}(r)$
should form a canonical chart whose canonical coordinates are
spatial scalars, and momenta are scalar densities. This determines
the form of the supermomentum by the requirement that $H_{r}(r)$
generate Diff${\,\rm I\!R}$.  The same requirement had already fixed
the form of the supermomentum in the original canonical variables.
These considerations show that
\begin{equation}
\Pi_{R} R' - \Lambda P_{\Lambda}' = \Pi_{M} M' + \Pi_{\sf R} {\sf
R}' \,.  \label{eq:supermomenta-eq}
\end{equation}
We already know how the new canonical variables $M, \Pi_{M}$ and
${\sf R}$ depend on the original canonical variables. Therefore, by
substituting the expressions (\ref{eq:canF}) - (\ref{eq:canM}) and
(\ref{eq:PiM}) - (\ref{eq:sfR}) into (\ref{eq:supermomenta-eq}) we
are able to determine $\Theta$. This gives us the missing
transformation equation for $\Pi_{\sf R}\,$:
\begin{eqnarray}
\lefteqn{\Pi_{\sf R} = \Pi_{R} - \frac{1}{2} R^{-1} \Lambda
P_{\Lambda} - \frac{1}{2} R^{-1} F^{-1} \Lambda P_{\Lambda}}
\nonumber \\ & & - R^{-1} \Lambda^{-2} F^{-1} \big(\, (\Lambda
P_{\Lambda})' (RR') - (\Lambda P_{\Lambda}) (RR')' \, \big) \,.
\label{eq:PisfR}
\end{eqnarray}
By inspecting the form of the constraints, we see that $\Pi_{\sf R}$
can be expressed as their linear combination:
\begin{equation}
\Pi_{\sf R} = F^{-1} \big(\, R^{-1}P_{\Lambda}\,H + R'
\Lambda^{-2}\, H_{r} \big) \,. \label{eq:PisfRconstr}
\end{equation}

     Our task is now clear: We must prove that the transition
\begin{equation}
\Lambda(r) , P_{\Lambda}(r);\; R(r), \Pi_{R}(r) \;\; \mapsto \;\;
M(r),\Pi_{M}(r);\; {\sf R}(r), \Pi_{\sf R}(r) \label{eq:can.trans}
\end{equation}
given by (\ref{eq:canF}) - (\ref{eq:canM}), (\ref{eq:PiM}) -
(\ref{eq:sfR}), and (\ref{eq:PisfR}) is a canonical transformation.
We prove this by showing that the difference of the Liouville forms
is an exact form:
\begin{eqnarray}
\lefteqn{\int_{-\infty}^{{\infty}} dr \, \big( P_{\Lambda}(r) \delta
\Lambda(r) + \Pi_{R}(r) \delta R (r)\, \big) } \nonumber \\ & &
-\int_{-\infty}^{\infty} dr \, \big(\, \Pi_{M}(r) \delta M(r) +
\Pi_{\sf R}(r) \delta {\sf R}(r)\, \big) = \delta \omega \big[
\Lambda, P_{\Lambda}, R, \Pi_{R} \big] \,.  \label{eq:Liouville}
\end{eqnarray}

     The difference of the integrands is evaluated by
straightforward rearrangements:
\begin{eqnarray}
\lefteqn{ P_{\Lambda} \delta \Lambda + \Pi_{R} \delta R - \Pi_{M}
\delta M -\Pi_{\sf R} \delta {\sf R} = } \nonumber \\ & & \delta
\left( \Lambda P_{\Lambda} + \frac{1}{2} RR' \ln \left| \frac{RR' -
\Lambda P_{\Lambda}}{RR' + \Lambda P_{\Lambda}} \right| \,\right) +
\nonumber \\ & & \left( \frac{1}{2}\, R \delta R \ln\left| \frac{RR'
+ \Lambda P_{\Lambda}} {RR' - \Lambda P_{\Lambda}} \right|\,
\right)' \label{eq:integrands}
\end{eqnarray}
To prove (\ref{eq:Liouville}), we integrate (\ref{eq:integrands}) in
$r$ and argue that the boundary terms
\begin{equation}
\frac{1}{2}\,R \delta R \ln \left| \frac{RR' + \Lambda P_{\Lambda}}
{RR' - \Lambda P_{\Lambda}} \right| \label{eq:Lio-boundary}
\end{equation}
vanish.

     At $r \rightarrow \pm \infty$, the falloff conditions imply
that $\Lambda \rightarrow 1$, $ R \rightarrow |r|$, $R' \rightarrow
\pm 1$, $P_{\Lambda} = O(|r|^{-\epsilon})$, and $\delta R =
O(|r|^{-\epsilon})$. The boundary term is of the order
\begin{equation}
 \frac{1}{2} \left[ R \delta R \ln \left|1 + \frac{2 \Lambda
P_{\Lambda}}{RR'}\right|\, \right] ^{\infty}_{-\infty} \approx \Big[
\delta R \left| R'^{-1} \Lambda P_{\Lambda}\right|\; \Big]
^{\infty}_{-\infty} \approx O(|r|^{-\epsilon})
\end{equation}
and hence vanishes at infinities.

     There are also boundary terms at the horizon, where $R \delta
R$ is finite, but the logarithm becomes infinite. However, due to
the absolute value within the logarithm, the infinite boundary term
inside the horizon matches the infinite boundary term outside the
horizon, and they can be considered as canceling each other when the
integral of the derivative term is interpreted through its principal
value. This proves (\ref{eq:Liouville}) and identifies $\omega \,$:
\begin{equation}
\omega [R, \Lambda, P_{\Lambda}] = \int_{-\infty}^{\infty} dr \,
\left( \Lambda P_{\Lambda} + \frac{1}{2} RR' \ln \left| \frac{RR' -
\Lambda P_{\Lambda}}{RR' + \Lambda P_{\Lambda}} \right| \, \right)
\,. \label{eq:omega}
\end{equation}

     The functional (\ref{eq:omega}) is well defined. The falloff of
the canonical variables at infinities implies
\begin{equation}
\frac{1}{2} RR' \ln \left| \frac{RR' - \Lambda P_{\Lambda}}{RR' +
\Lambda P_{\Lambda}} \right| \approx - \Lambda P_{\Lambda} +
O(|r|^{-(1+\epsilon)}) \,.
\end{equation}
The integrand of (\ref{eq:omega}) thus falls faster than $|r|^{-1}$,
which avoids the logarithmic singularity. Close to the horizon
$r=r_{0}$, the integrand of (\ref{eq:omega}) behaves as $\ln |r -
r_{0}|$, and hence the integral from a given $r$ to $r_{0}$ stays
finite.

     Equations (\ref{eq:Liouville}) and (\ref{eq:omega}) lead to the
generating functional of the canonical transformation
(\ref{eq:can.trans}). The generating functional $ \Omega
[P_{\Lambda}, \Pi_{R};\,M,{\sf R}]$ emerges when we introduce the
{\em old momenta} and {\em new coordinates} as a new coordinate
chart on the phase space, rewrite (\ref{eq:Liouville}) in the form
\begin{eqnarray}
\lefteqn{-\,\int_{-\infty}^{\infty} dr \big( \Lambda (r) \delta
P_{\Lambda}(r) + R(r) \delta R(r) \big)} \nonumber \\ & & -
\int_{-\infty}^{\infty} dr \, \big( \Pi_{M}(r) \delta M(r) +
\Pi_{\sf R}(r) \delta {\sf R}(r) \big) = \delta \Omega [P_{\Lambda},
\Pi_{R};\,M,{\sf R}] \,, \label{eq:LiouvilleOmega}
\end{eqnarray}
and express
\begin{eqnarray}
\lefteqn{\Omega \; \mbox{$:=$} \; -\int^{\infty}_{-\infty} dr \,
\big( \Lambda(r) P_{\Lambda}(r) + R(r) \Pi_{R}(r) \, \big) + \omega
[\Lambda, P_{\Lambda}, R, \Pi_{R}]} \nonumber \\ &
&=\int^{\infty}_{-\infty} dr \left( - R \Pi_{R} + \frac{1}{2} RR'
\ln \left| \frac{RR' - \Lambda P_{\Lambda}}{RR' + \Lambda
P_{\Lambda}}\right|\, \right) \label{eq:defOmega}
\end{eqnarray}
in the new chart. This is done by calculating $\Lambda > 0$ from the
mass equation (\ref{eq:canM}),
\begin{equation}
\Lambda = \frac{|RR'|}{\sqrt{R(R-2M) + P_{\Lambda}{}^{2}}} \,,
\end{equation}
and substituting it back into (\ref{eq:defOmega}). The result can be
written in the form
\begin{eqnarray}
\lefteqn{\Omega [P_{\Lambda}, \Pi_{R};\,M,{\sf R}] =
-\int^{\infty}_{-\infty} dr \, {\sf R}(r) \Pi_{R}(r)\; +} \nonumber
\\ & &\int_{-\infty}^{\infty} dr\, \frac{1}{2} {\sf RR}' \ln \left|
\frac{\sqrt{{\sf R(R} - 2M) + P_{\Lambda}{}^{2}} - P_{\Lambda}}
{\sqrt{{\sf R(R} - 2M) + P_{\Lambda}{}^{2}} + P_{\Lambda}} \right|
\,.  \label{eq:Omega}
\end{eqnarray}
By comparing the coefficients of the independent variations $\delta
P_{\Lambda}, \delta \Pi_{R}, \delta M, \delta {\sf R}$ in
(\ref{eq:LiouvilleOmega}), we generate the canonical transformation
(\ref{eq:can.trans}) by $\Omega\,$:
\begin{equation}
\begin{array}{ll}
\Lambda(r) = -\,\delta \Omega / \delta P_{\Lambda}(r)\,,& \Pi_{M}(r)
= -\,\delta \Omega / \delta M(r)\,, \\ R(r) = -\,\delta \Omega /
\delta \Pi_{R}(r)\,,& \Pi_{\sf R}(r) = -\,\delta \Omega / \delta
{\sf R}(r)\,.
\end{array}
\end{equation}
When resolved with respect to the new canonical variables, these
equations give our old transformation equations (\ref{eq:canM}),
(\ref{eq:PiM}), (\ref{eq:sfR}), and (\ref{eq:PisfR}).

One should perhaps note that both integrands of (\ref{eq:Omega})
fall at infinities at a slow rate, as $O(|r|^{-\epsilon})$, and
hence the generating functional $\Omega$ is singular.  This happens
because we added to the well-defined functional $\omega$ an
ill-defined term
\begin{equation}
-\, \int^{\infty}_{-\infty} dr \, \big( P_{\Lambda}\Lambda +
\Pi_{R}R \, \big)\,.
\end{equation}
The simplest way out of this difficulty is to let the momenta fall
off faster than $O(|r|^{-\epsilon})$. We have seen this can always
be done for the Schwarzschild black hole, (\ref{eq:strongfall}).

     From the transformation formulae (\ref{eq:canF}) -
(\ref{eq:canM}), (\ref{eq:PiM}) - (\ref{eq:sfR}), (\ref{eq:PisfR}),
and the {\em weak} falloff conditions (\ref{eq:Lambdafall}) -
(\ref{eq:PiRfall}) we easily deduce the falloff of the new canonical
variables:
\begin{eqnarray}
M(t,r)&=& M_{\pm}(t) + O^{\infty}(|r|^{-\epsilon})\,,
\label{eq:Mfall}\\ {\sf R}(t,r)&=&|r| +
O^{\infty}(|r|^{-\epsilon})\,, \label{eq:sfRfall} \\
\Pi_{M}(t,r)&=&O^{\infty}(|r|^{-(1+\epsilon)})\,, \label{eq:PiMfall}
\\ \Pi_{\sf R}(t,r)&=&O^{\infty}(|r|^{-(1+\epsilon)})
\label{eq:PisfRfall}\,.
\end{eqnarray}
Recalling that $\Pi_{M}(r) = -T'(r)$, we can integrate
(\ref{eq:PiMfall}) with respect to $r$ and prove the earlier
statement (\ref{eq:Tfall}) about the behavior of the foliation at
infinities.

     The old canonical variables are continuous (and sufficiently
differentiable) functions of $r$ even across the horizon. The
transformation equations imply that the new canonical coordinates
$M(r)$ and ${\sf R}(r)$ are also continuous across the horizon, but
this cannot be said about their conjugate momenta.  Equations
(\ref{eq:PiM}) and (\ref{eq:PisfRconstr}) indicate that $\Pi_{M}$
and $\Pi_{\sf R}$ are both proportional to $F^{-1}$. While the
coefficients of $F^{-1}$ are continuous, $F$ goes to zero on the
horizon, and generically changes its sign. As a result, $\Pi_{M}$
and $\Pi_{\sf R}$ become infinite on the horizon, and generically
suffer an infinite jump.

     This means that when the canonical data are such that $F$
vanishes for some $r$ (which, in particular, always happens for the
Schwarzschild solution), our canonical transformation becomes
singular. In other words, $\omega$ is not a differentiable
functional of the old canonical variables. One must use the new
canonical variables with caution.

     Except at the horizon, the canonical transformation
(\ref{eq:can.trans}) can be inverted for the old variables:
\begin{eqnarray}
\Lambda &=& \big( {\sf F}^{-1} {\sf R}'^{2} - {\sf F} \Pi_{M}{}^{2}
\big) ^{1/2} \,, \label{eq:Lambdanew} \\ P_{\Lambda} &=& {\sf R}
{\sf F} \Pi_{M} \, \big( {\sf F}^{-1} {\sf R}'^{2} - {\sf F}
\Pi_{M}{}^{2} \big) ^{-1/2} \,, \label{eq:PiLambdanew}\\ R & =& {\sf
R} \,, \\ \Pi_{R} &=& \Pi_{\sf R} + {\frac{1}{2}} \Pi_{M} \nonumber
\\ & & + R^{-1}{\sf F}^{-1}\big({\sf F}^{-1} {\sf R}'^{2} - {\sf F}
\Pi_{M}{}^{2} \big) ^{-1} \big( ( {\sf R} {\sf F} \Pi_{M} )' ({\sf
RR}') - ( {\sf R} {\sf F} \Pi_{M} )( {\sf RR}')' \big) \nonumber \\
& & + \frac{1}{2} {\sf F} \Pi_{M} \,.
\end{eqnarray}
In these equations, ${\sf F}$ is an abbreviation for
\begin{equation}
{\sf F} = 1 - 2M{\sf R}^{-1} \,.  \label{eq:FofM}
\end{equation}

\section{$M(t,r)$ As a Constant of Motion}

     Guided by the spacetime form of the Schwarzschild solution, we
have introduced the Schwarzschild mass $M(r)$ as a dynamical
variable on our phase space. We shall now prove that if the
canonical data satisfy the constraints, the mass function $M(r)$
does not depend on $r$, and if they also satisfy the Hamilton
equations, $M(r)$ is a constant of motion.

     Both statements follow by straightforward algebra. By
differentiating the definition (\ref{eq:canM}) of $M(r)$ with
respect to $r$, we find that $M'(r)$ is a linear combination of the
constraints (\ref{eq:SH}) and (\ref{eq:SM}):
\begin{equation}
M' = - \Lambda^{-1} \big(\, R' \, H + R^{-1} P_{\Lambda} \,
H_{r}\,\big) \,.  \label{eq:M'}
\end{equation}
Because $M(r)$ is a spatial scalar,
\begin{equation}
\big\{ M(r), H_{r}(r') \big\} = M'(r)\, \delta (r, r') \,,
\end{equation}
equation (\ref{eq:M'}) can be translated into the statement that the
Poisson bracket of $M(r)$ with the supermomentum weakly vanishes.

     Equations (\ref{eq:M'}) and (\ref{eq:PisfRconstr}) express $M'$
and $\Pi_{\sf R}$ as linear combinations of the constraints.
Inversely, we can express the constraints in terms of the new
canonical variables.  We already know that
\begin{equation}
H_{r} = \Pi_{\sf R} {\sf R}' + \Pi_{M} M' \,. \label{eq:Hrnewvar}
\end{equation}
The super-Hamiltonian $H$ is then calculated from (\ref{eq:M'}). By
using (\ref{eq:Lambdanew}) and (\ref{eq:PiLambdanew}) we obtain
\begin{equation}
H = - \, \frac{{\sf F}^{-1}M' {\sf R}' + {\sf F} \Pi_{M} \Pi_{\sf
R}} { \big( {\sf F}^{-1} {\sf R}'^{2} - {\sf F} \Pi_{M}{}^{2} \big)
^{1/2}} \,.  \label{eq:Hnew}
\end{equation}

    These expressions are useful for showing the closure of the
Poisson bracket
\begin{equation}
\big\{ M(r), H(r') \big\} = - \Lambda^{-3} R' \, H_{r}\, \delta
(r,r') \,.  \label{eq:MHbracket}
\end{equation}
The same calculation in terms of the old variables is much more
cumbersome.  One should note that the right-hand side of
(\ref{eq:MHbracket}) does not contain the super-Hamiltonian, only
the supermomentum. Because $H(r)$ generates the time evolution,
(\ref{eq:MHbracket}) means that $M(r)$ is a constant of motion.

     From equations (\ref{eq:Hrnewvar}) - (\ref{eq:Hnew}) we can
conclude that the Hamiltonian and momentum constraints
\begin{equation}
H(r) = 0 \,, \;\;\; H_{r}(r) = 0 \label{eq:Hconstraints}
\end{equation}
are entirely equivalent to a new set of constraints,
\begin{equation}
M'(r) = 0 \,, \;\;\; \Pi_{\sf R}(r) = 0, \label{eq:newconstraints}
\end{equation}
except on the horizon.

     On the horizon, we need to be more circumspect. Let the old
canonical variables satisfy the constraints (\ref{eq:Hconstraints}).
We know that such canonical data correspond to the Schwarzschild
solution, and hence there are at most two values of $r$ for which
$F=0$.  From (\ref{eq:M'}) and (\ref{eq:PisfRconstr}) we conclude
that $M'(r)=0$ everywhere, and $\Pi_{R}(r)=0$ except at the horizon
points.  If we insist that $\Pi_{R}(r)$ be a continuous function of
$r$ when the data satisfy the constraints, we conclude that
$\Pi_{R}(r)=0$ everywhere.

     Inversely, let us impose the constraints
(\ref{eq:newconstraints}) on the new canonical variables. Except at
the horizon points, equations (\ref{eq:Hrnewvar}) and
(\ref{eq:Hnew}) imply the old constraints (\ref{eq:Hconstraints}).
At the horizon points, $F(r)=0$, and $\Pi_{M}(r)$ becomes infinite
in such a way that $F(r)P_{M}(r)$ stays finite (cf.~(\ref{eq:PiM})).
We can again argue from continuity that the new constraints imply
the old constraints even at the horizon points. In this sense, the
constraint systems (\ref{eq:Hconstraints}) and
(\ref{eq:newconstraints}) are equivalent everywhere.

\section{The Tale of Three Actions} \label{sec-tale}

     Written in terms of the new canonical variables $M, \Pi_{M},
{\sf R}$ and $\Pi_{\sf R}$, the hypersurface action becomes
\begin{eqnarray}
\lefteqn{S_{\Sigma} \big[ M, \Pi_{M}, {\sf R}, \Pi_{\sf R};\, N,
N^{r} \big] = } \nonumber \\ & & \int dt \int^{\infty}_{-\infty} dr
\, \Big(\, \Pi_{M}(r) \dot{M}(r) + \Pi_{\sf R}(r) \dot{\sf R}(r)
\nonumber \\ & & - N(r) H(r) - N^{r} H_{r}(r) \, \Big) \,.
\label{eq:newSSigma}
\end{eqnarray}
The super-Hamiltonian $H$ and supermomentum $H_{r}$ are now
functions (\ref{eq:Hnew}) and (\ref{eq:Hrnewvar}) of the new
variables.  The variation of (\ref{eq:newSSigma}) with respect to
$N$ and $N^{r}$ imposes the Hamiltonian and momentum constraints
(\ref{eq:Hconstraints}).  We have found that these constraints are
equivalent to a new set of constraints, (\ref{eq:newconstraints}),
which are simple functions of the new variables.  The action
(\ref{eq:newSSigma}) is equivalent to a new action
\begin{eqnarray}
\lefteqn{S_{\Sigma} \big[ M, \Pi_{M}, {\sf R}, \Pi_{\sf R};\, N^{M},
N^{\sf R} \big] =} \nonumber \\ & & \int dt \int^{\infty}_{-\infty}
dr \, \big(\, \Pi_{M}(r) \dot{M}(r) + \Pi_{\sf R}(r) \dot{\sf
R}(r)\, \big) \nonumber \\ & & -\, \int dt \int^{\infty}_{-\infty}
dr \, \big(\, N^{M}(r) M'(r) + N^{\sf R}(r) \Pi_{\sf R}(r) \, \big)
\,, \label{eq:Snewconstr}
\end{eqnarray}
in which the new constraints, rather than the old ones, are adjoined
to the Liouville form. This is done by a new set, $ N^{M}(r)$ and $
N^{\sf R}(r)$, of Lagrange multipliers.  The falloff conditions
(\ref{eq:Mfall}) - (\ref{eq:PisfRfall}) imply that the
super-Hamiltonian almost coincides with $M'$ at infinities:
\begin{equation}
H(r) = \mp M'(r) + O^{\infty}(|r|^{-(2+\epsilon)}) \,.
\end{equation}
The asymptotic values of the multipliers $N$ and $N^{M}$ are thus
related by
\begin{equation}
N^{M}_{\pm}(t) = \mp N_{\pm}(t) \,.
\end{equation}

     The hypersurface action must again be complemented by the ADM
boundary action (\ref{eq:Sbound}). In the new variables, the ADM
energy (\ref{eq:ADMenergy}) is the value of the mass function $M(r)$
at infinity, $E_{\pm} = M_{\pm}$.  The boundary action, like the
hypersurface action (\ref{eq:Snewconstr}), is again a very simple
function of the new variables:
\begin{equation}
S_{\partial \Sigma} = - \, \int dt \, \big( N_{+}M_{+} + N_{-}M_{-}
\big) \,.
\end{equation}
The total action is the sum
\begin{equation}
S \big[ M, \Pi_{M}, {\sf R}, \Pi_{\sf R};\, N^{M}, N^{\sf R} \big] =
S_{\Sigma} \big[ M, \Pi_{M}, {\sf R}, \Pi_{\sf R};\, N^{M}, N^{\sf
R} \big] + S_{\partial \Sigma} \big[ M;\,N^{M} \big] \,.
\label{eq:Sunpar}
\end{equation}
It is transparent how the boundary action cancels the boundary term
obtained by varying $M(r)$ in the hypersurface action.

     The lapse functions $N_{\pm}(t)$ at infinities must be treated
as prescribed functions of the time parameter $t$. After the
constraints are imposed, the boundary part of the total action
survives as a true, $t$-dependent, Hamiltonian of the black hole. We
shall discuss this reduction process in the next section.

     The ends $N_{\pm}(t)$ of $N(t,r)$ are freed by parametrizing
the action at infinities:
\begin{equation}
S_{\partial \Sigma}\big[ M;\,\tau_{+}, \tau_{-} \big] = - \, \int dt
\, \big( M_{+} \dot{\tau}_{+} - M_{-} \dot{\tau}_{-} \big) \,.
\label{eq:parboundS}
\end{equation}
The total action
\begin{eqnarray}
\lefteqn {S \big[ M, \Pi_{M}, {\sf R}, \Pi_{\sf R};\, \tau_{+},
\tau_{-};\, N^{M}, N^{\sf R} \big] = } \nonumber \\ & & S_{\Sigma}
\big[ M, \Pi_{M}, {\sf R}, \Pi_{\sf R}; \, N^{M}, N^{\sf R} \big] +
S_{\partial \Sigma}\big[ M;\,\tau_{+}, \tau_{-} \big]
\label{eq:Shamlag}
\end{eqnarray}
depends now on two additional variables, $\tau_{\pm}(t)$, which can
also be freely varied. It is no longer a canonical action, because
$\tau_{+}$ and $\tau_{-}$ do not come with their conjugate momenta.

     There are two entirely different ways in which the action
(\ref{eq:Shamlag}) can be brought into canonical form. The first one
is standard, the second one rather unexpected. Let us explain the
standard method first.

     One can say that the action (\ref{eq:Shamlag}) has a mixed
Hamiltonian - Lagrangian form, and that $\tau_{\pm}(t)$ is a pair of
Lagrangian configuration variables. The purely Hamiltonian form
should be reached by the Legendre dual transformation which
complements $\tau_{\pm}$ by the conjugate momenta $\pi_{\pm}$.
However, because the action is linear in the velocities
$\dot{\tau}_{\pm}$, as soon as one starts implementing this program,
one obtains two new constraints
\begin{equation}
\begin{array}{lcl}
C_{+} \; & \mbox{$:=$} & \; \pi_{+} + M_{+} = 0 \,, \nonumber \\
C_{-} \; & \mbox{$:=$} & \; - \pi_{-} + M_{-} = 0 \,. \label{eq:C}
\end{array}
\end{equation}
They have vanishing Poisson brackets among themselves, and with the
hypersurface constraints $M'(r) = 0 = \Pi_{\sf R}(r)$. The resulting
constraint system is thus first class. The new constraints $C_{\pm}$
must be adjoined to the canonical action by Lagrange multipliers
$N_{\pm}\,$:
\begin{eqnarray}
\lefteqn{ S \big[ M, \Pi_{M}, {\sf R}, \Pi_{\sf R};\, \tau_{+},
\pi_{+}, \tau_{-}, \pi_{-};\, N^{M}, N^{\sf R}, N_{+}, N_{-} \big] =
} \nonumber \\ & & \int dt \int^{\infty}_{\infty} dr \, \Big( \,
\Pi_{M}(r) \dot{M}(r) + \Pi_{\sf R}(r) \dot{\sf R}(r) - N^{M}(r)
M'(r) - N^{\sf R}(r) \Pi_{\sf R}(r) \, \Big) \nonumber \\ & & +\,
\int dt \, \Big( \pi_{+} \dot{\tau}_{+} + \pi_{-} \dot{\tau}_{-} -
N_{+}C_{+} - N_{-}C_{-} \Big) \,.  \label{eq:Scanpi}
\end{eqnarray}
The variation of (\ref{eq:Scanpi}) with respect to the momenta
$\pi_{\pm}$ leads back to equation (\ref{eq:Npar}). The new
multipliers $N_{\pm}$ thus are what the symbols suggest: the lapse
function at infinities.

     The price we paid for the canonical form (\ref{eq:Scanpi}) was
a couple of new variables and a couple of new constraints. It is
gratifying to learn that one can get the same product for free: The
mixed variables $M, \Pi_{M};\, \tau_{+}, \tau_{-}$ in the action
(\ref{eq:Shamlag}) can simply be transformed into a canonical chart.
It is even more gratifying that the new canonical variables have a
clean geometric meaning: they turn out to be the Killing time $T(r)$
and the mass density $P_{T}(r)$ along the hypersurface, complemented
by a canonical pair of constants of motion.

     To see how this comes about, notice that the action
(\ref{eq:Shamlag}) is linear in the time derivatives $\dot{M} (r),
\dot{\tau}_{\pm}$. The homogeneous part of (\ref{eq:Shamlag}) thus
defines a one-form
\begin{equation}
\Theta \; \mbox{$:=$} \; \int_{-\infty}^{\infty} dr \, \Pi_{M}(r)
\delta M(r) - \big( M_{+} \delta \tau_{+} - M_{-} \delta \tau_{-}
\big) \label{eq:Thetis}
\end{equation}
on $\big( M(r), \Pi_{M}(r);\, \tau_{+}, \tau_{-}\big)\,$.

     We shall now cast (\ref{eq:Thetis}) into a Liouville form.
First, we replace the mass function $ M(r)$ by the mass at left
infinity $m$, and by the mass density ${\Gamma }(r)$:
\begin{equation}
m = M_{-}, \;\;\; \Gamma (r) = M'(r) \,.
\end{equation}
Inversely,
\begin{equation}
M(r) = m + \int_{-\infty}^{r} dr' \, \Gamma(r') \,.
\label{eq:massdecomp}
\end{equation}
By introducing (\ref{eq:massdecomp}) into (\ref{eq:Thetis}) we get
\begin{eqnarray}
\lefteqn{ \Theta = \left( \big( \tau_{+} - \tau_{-} \big) +
\int_{-\infty}^{\infty} dr \, \Pi_{M}(r) \right) \, \delta m }
\nonumber \\ & & +\,\int_{-\infty}^{\infty} dr \, \left( \tau_{+}
\delta \Gamma (r) + \Pi_{M}(r) \int_{-\infty}^{r} dr' \, \delta
\Gamma (r') \right) \nonumber \\ & & +\, \delta \big( M_{-} \tau_{-}
- M_{+} \tau_{+} \big) \,.  \label{eq:Thetis2}
\end{eqnarray}
To rearrange (\ref{eq:Thetis2}), we write the identity
\begin{eqnarray}
\lefteqn{ \left( \int_{\infty}^{r} dr'\, \Pi_{M}(r') \, \times
\int_{-\infty}^{r} dr'\, \delta \Gamma(r') \right)' \, = } \nonumber
\\ & & \Pi_{M}(r) \int_{-\infty}^{r} dr' \, \delta \Gamma (r') +
\delta \Gamma (r) \, \int_{\infty}^{r} dr' \, \Pi_{M}(r') \,,
\end{eqnarray}
which we then integrate from $r=-\infty$ to $r=\infty$:
\begin{equation}
 \int_{-\infty}^{\infty} dr\, \Pi_{M}(r) \int_{-\infty}^{r} dr' \,
\delta \Gamma (r') = - \, \int_{-\infty}^{\infty} dr \, \delta
\Gamma (r) \int_{\infty}^{r} dr' \, \Pi_{M}(r') \,.
\label{eq:auxid}
\end{equation}
We immediately see that
\begin{eqnarray}
\Theta &=& \left( \big( \tau_{+} - \tau_{-} \big) +
\int_{-\infty}^{\infty} dr' \, \Pi_{M}(r') \right) \delta m
\nonumber \\ & & +\, \int_{-\infty}^{\infty} dr \, \left( \tau_{+} -
\int_{\infty}^{r} dr' \, \Pi_{M}(r') \right) \delta \Gamma (r)
\nonumber \\ & & +\;\delta \big( M_{-} \tau_{-} - M_{+} \tau_{+}
\big) \,.
\end{eqnarray}
This shows that
\begin{eqnarray}
m &=& M_{-} \,, \label{eq:m} \\ p &=& \big( \tau_{+} - \tau_{-}
\big) + \int_{-\infty}^{\infty} dr' \, \Pi_{M}(r') \label{eq:p}
\end{eqnarray}
and
\begin{eqnarray}
\Gamma (r) &=& M'(r) \,, \label{eq:Gamma} \\ \Pi_{\Gamma}(r) &=&
\tau_{+} - \int_{\infty}^{r} dr' \, \Pi_{M}(r') \label{eq:PiGamma}
\end{eqnarray}
is a canonical chart: Indeed,
\begin{eqnarray}
\Theta &=& p \delta m + \int_{-\infty}^{\infty} dr \,
\Pi_{\Gamma}(r) \delta \Gamma (r) \\ & & +\; \delta \big( M_{-}
\tau_{-} - M_{+} \tau_{+} \big) \,.  \label{ThetaLiou}
\end{eqnarray}
differs from the Liouville form only by an exact form.

     When passing from $\Pi_{M}(r) = - T'(r)$ to $T(r)$ we had to
fix a constant of integration. The Killing time (\ref{eq:PiGamma})
is fixed by requiring that it matches the proper time $\tau_{+}$ on
the parametrization clock at right infinity. By construction, $T(r)$
is a spatial scalar and $\Gamma (r)$ a scalar density. We would like
to change $T(r)$ into a canonical coordinate, and $\Gamma (r)$ into
a canonical momentum. This is done by an elementary canonical
transformation (in the sense of Carath\'{e}odory
\cite{Caratheodory})
\begin{eqnarray}
T(r) & = & \Pi_{\Gamma}(r)\; =\; \tau_{+} - \int_{\infty}^{r} dr' \,
\Pi_{M}(r') \,, \label{eq:Killingtime} \\ \Pi_{T}(r) &=& - \Gamma
(r)\; =\; - M'(r) \label{eq:PiT}
\end{eqnarray}
which sends $\Theta$ into
\begin{equation}
\Theta = p \delta m + \int_{-\infty}^{\infty} dr \, \Pi_{T}(r)
\delta T(r) +\; \delta \omega \,, \label{ThetaLioufin}
\end{equation}
with
\begin{eqnarray}
 \omega &=& \big( M_{-} \tau_{-} - M_{+} \tau_{+} \big) +
\int_{-\infty}^{\infty} dr \, P_{\Gamma}(r) \Gamma(r) \nonumber \\ &
= & M_{-} \big( \tau_{-} - \tau_{+} \big) - \int_{-\infty}^{\infty}
dr \, \int_{\infty}^{r} dr' \, \Pi_{M}(r') M'(r') \,.
\label{eq:omega2}
\end{eqnarray}
The last equations (\ref{ThetaLioufin}), (\ref{eq:omega2}) show that
the transformation
\begin{equation}
\tau_{+}, \tau_{-};\,M(r), \Pi_{M}(r) \;\; \mapsto \;\; m,p; \,
T(r), \Pi_{T}(r) \label{eq:canchart}
\end{equation}
given by equations (\ref{eq:m}) - (\ref{eq:p}) and
(\ref{eq:Killingtime}) - (\ref{eq:PiT}) constructs a canonical chart
from the originally mixed variables.

      The transformation (\ref{eq:canchart}) casts the parametrized
action (\ref{eq:Snewconstr}), (\ref{eq:parboundS}) -
(\ref{eq:Shamlag}) to an extremely simple canonical form
\begin{eqnarray}
\lefteqn{S \big[m, p;\, T, \Pi_{T}, {\sf R}, \Pi_{\sf R};\, N^{T},
N^{\sf R} \big] = } \nonumber \\ & & \int dt \left( p \dot{m} +
\int^{\infty}_{-\infty} dr \, \big(\, \Pi_{T}(r) \dot{T}(r) +
\Pi_{\sf R}(r) \dot{\sf R}(r)\, \big) \right) \nonumber \\ & & -
\int dt \int^{\infty}_{-\infty} dr \, \big(\, N^{T}(r) \Pi_{T}(r) +
N^{\sf R}(r) \Pi_{\sf R}(r) \, \big) \,, \label{eq:Taction}
\end{eqnarray}
(We gave the multiplier $-N^{M}(r)$ a new name $N^{T}(r)$.)  In the
transition from (\ref{eq:Shamlag}) to (\ref{eq:Taction}) we have
discarded the total time derivative $\dot{\omega}$. Such a procedure
does not change equations of motion, and it is used throughout
classical mechanics. When applied to canonical action, it generates
canonical transformations. Here we have used it for bringing the
action to canonical form.

     Because the multipliers $N^{T}(r)$ and $N^{\sf R}(r)$ are
freely variable, the action (\ref{eq:Taction}) enforces the
constraints
\begin{equation}
\Pi_{T}(r) = 0\,, \;\;\; \Pi_{\sf R}(r)=0 \,. \label{PiTconstr}
\end{equation}
The boundary action disappeared from (\ref{eq:Taction}) and the
Hamiltonian took the form
\begin{equation}
H[N^{T}] + H[N^{\sf R}] \;\mbox{$:=$}\; \int_{-\infty}^{\infty} dr\,
N^{T}(r)\Pi_{T}(r) + \int_{-\infty}^{\infty} dr\, N^{\sf
R}(r)\Pi_{\sf R}(r) \,. \label{eq:hamNT}
\end{equation}
It is a linear combination of constraints, and as such it weakly
vanishes.

     It is well known that the smeared super-Hamiltonian
\begin{equation}
H[N] = \int_{-\infty}^{\infty} dr\, N(r)H(r)
\end{equation}
in the Dirac-ADM action (\ref{eq:S-ham}) generates the change of the
canonical data when the hypersurface is displaced by the proper time
$N(r)$ in the normal direction. Similarly, the smeared supermomentum
\begin{equation}
H_{r}[N^{r}] = \int_{-\infty}^{\infty} dr\, N^{r}(r)H_{r}(r)
\end{equation}
generates the change of the data when the hypersurface is shifted by
the tangential vector $N^{r}(r)$.

     The Hamiltonian (\ref{eq:hamNT}) generates a different type of
displacement. The Hamilton equations
\begin{equation}
\begin{array}{lclcl}
\dot{T}(t,r) &=& \big \{ T(t,r),\,H[N^{T}] \big \} &=&
N^{T}(t,r)\,,\\ \dot{\sf R}(t,r) &=& \big \{ {\sf R}(t,r),\,H[N^{T}]
\big \}& =& 0
\end{array}
\end{equation}
reveal how $H[N^{T}]$ displaces the hypersurface in the Kruskal
diagram. It shifts it along the lines of constant $\sf R$ by the
amount $N^{T}(t,r)$ of Killing time which differs from one line to
another. Similarly, $H[N^{\sf R}]$ displaces the hypersurface along
the lines of a constant $T$ in such a way that the curvature
coordinate $\sf R$ changes by the amount $N^{\sf R}(t,r)$. The
Hamiltonian (\ref{eq:hamNT}) thus generates spacetime
diffeomorphisms in $\big( T, {\sf R} \big)$. The elaboration of this
general point can be found in Isham and Kucha\v{r}
\cite{Isham-Kuchar}.

     The new constraints (\ref{PiTconstr}) have a very simple form:
a number of canonical momenta is set equal to zero. Locally, any
system of first-class constraints can be brought to such a form, but
it is usually impossible to find explicitly the necessary
transformation. Our result demonstrates that this is feasible for
Schwarzschild black holes, and that it can be done globally. The
momenta which are required to vanish are conjugate to the embedding
variables $T(r)$ and ${\sf R}(r)$ which locate the hypersurface in
the Kruskal diagram.

     There is a pair of canonically conjugate variables in the
action (\ref{eq:Taction}), namely, $m(t)$ and $p(t)$, which is not
subject to any constraints. However, there is no nonvanishing
Hamiltonian in (\ref{eq:Taction}) which would evolve these
variables. Both $m$ and $p$ are thus constants of motion. The
meaning of $m$ as the mass at infinity is clear, and so is its
conservation. The significance of $p$ is at first puzzling. The
momentum $p$ was introduced by the transformation (\ref{eq:p}). Due
to (\ref{eq:Killingtime}), $p$ can be interpreted as
\begin{equation}
p(t) = T_{-}(t) - \tau_{-}(t) \,.  \label{eq:Ttau}
\end{equation}
But should not the Killing time coincide with the parametrization
time also at left infinity, and should not thus $p$ simply vanish?
Are we not missing a constraint?

     The answer to this question is {\em no\,}. The times
$\tau_{\pm}$ are introduced by the parametrization process and their
{\em origins} are entirely independent. The left infinity does not
know what the right infinity does. When we shift the origins by {\em
different amounts} $\,\alpha _{-} = const$ and $\alpha _{+} =
const\,$ at the left and right infinities,
\begin{equation}
\tau_{-}(t) \mapsto \tau_{-}(t) + \alpha _{-} \,, \;\;\; \tau_{+}(t)
\mapsto \tau_{+}(t) + \alpha _{+}\,,
\end {equation}
the parametrized action is unchanged.

     The origin of the Killing time $T(r)$ is also arbitrary. We
have chosen it so that $T(r)$ matches $\tau_{+}$ at {\em right
infinity\,}.  Once defined this way, $T(r)$ can be used to propagate
the choice of the origin from right infinity to left infinity.
(After the constraints are imposed, and the Schwarzschild solution
is found, this propagation amounts to drawing the straight line
across the Kruskal diagram, from the $\tau_{+} (t) = 0$ point at
right infinity, through the bifurcation point of the horizon, and
all the way up to the left infinity.) There is, however, no reason
why the parametrization clock at left infinity should have been set
to zero at this propagated origin. The variable $p$ tells us the
difference between the origins of the parametrization times at the
right and left infinities. More precisely, $p$ is the value of the
Killing time $T(r)$ (which is matched to the parametrization time at
right infinity) at the origin of the parametrization time at left
infinity.  Once set, the parametrization clock and the Killing time
clock run at the same pace, both of them measuring intervals of
proper time.  Therefore, it does not matter {\em when} we read their
difference.  This is the reason why $p(t)$ of equation
(\ref{eq:Ttau}) is a constant of motion.

     To summarize, we have arrived at three different canonical
actions describing the same physical system, namely, primordial
black holes.  The {\em unparametrized canonical action}
(\ref{eq:Sunpar}) has a nonvanishing Hamiltonian. The two {\em
parametrized canonical actions} that follow have only constraints.
The first of these, (\ref{eq:Scanpi}), has more variables and more
constraints than are really necessary. Also, it does not disentangle
the variables which are constrained to vanish from those that
survive as true dynamical degrees of freedom. The last of our
actions, (\ref{eq:Taction}), sticks to the original number of
variables and constraints, and at the same time clearly identifies
the true degrees of freedom. We believe it provides the simplest
canonical framework for studying Schwarzschild black holes.

\section{Reduced Canonical Theory}

Each canonical action we have introduced predicts a Hamiltonian
evolution. To compare these evolutions, we first reduce the actions
to the same set of true degrees of freedom. The action is reduced by
solving the constraints and substituting the solutions back into the
action.

     Let us start with the unparametrized action (\ref{eq:Sunpar}).
The constraint $M'(r)=0$ tells us that only the homogeneous mode of
$M(r)$ survives:
\begin{equation}
M(t,r) = {\bf m}(t) \,.  \label{eq:m=M}
\end{equation}
By substituting (\ref{eq:m=M}) and $\Pi _{\sf R}(r) = 0$ back into
(\ref{eq:Sunpar}) we obtain
\begin{equation}
S \big[{\bf m}, {\bf p} \big] = \int dt \, \Big( {\bf p} \dot{\bf m}
- \big( N_{+}(t) + N_{-}(t) \big) {\bf m} \Big) \,.
\label{eq:redSnp}
\end{equation}
The form of the reduced action enabled us to identify
\begin{equation}
{\bf p} \;\mbox{$:=$}\; \int_{-\infty}^{\infty}dr\, \Pi_{M}(r)
\label{eq:bfp}
\end{equation}
as the momentum canonically conjugate to $\bf m$.  The reduced
action has one degree of freedom, $\bf m$, and a true time-dependent
Hamiltonian
\begin{equation}
{\bf h}(t,{\bf m,p}) = {\cal N}(t) {\bf m}\,, \;\;\; {\cal N}(t)
\;\mbox{$:=$}\; N_{+}(t) + N_{-}(t) \,.  \label{eq:trueh}
\end{equation}
This Hamiltonian is proportional to $\bf m$, with a coefficient
${\cal N}(t)$ which is a prescribed function of $t$. The Hamilton
equations of motion
\begin{equation}
\dot {\bf m} = \partial {\bf h}(t,{\bf m,p}) / \partial {\bf p} = 0
\,, \;\;\; \dot {\bf p} = - \partial {\bf h}(t,{\bf m,p)} / \partial
{\bf m} = - {\cal N}(t) \label{eq:motion}
\end{equation}
indicate that ${\bf m}(t)$ is a constant of motion, but ${\bf p}(t)$
changes in time.  This is consistent with (\ref{eq:bfp}) which
identifies ${\bf p}$ with $-(T_{+}-T_{-})$. The difference of the
Killing times between the left and the right infinities stays the
same only if we evolve the hypersurface by the lapse function which
has opposite values at $\pm \infty$, i.e.\, by ${\cal N}=0$.

     Next, reduce the parametrized action (\ref{eq:Scanpi}) by
solving both the hypersurface constraints and the additional
constraints (\ref{eq:C}). We get
\begin{equation}
S \big[{\bf m, p}, {\cal T} \big] = \int dt \, \big( {\bf p \dot{m}}
- \dot{\cal T} {\bf m} \big) \,, \label{eq:redparS}
\end{equation}
with
\begin{equation}
{\cal T} \;\mbox{$:=$}\;\; \tau_{+} - \tau_{-} \,.  \label{eq:calT}
\end{equation}
The action (\ref{eq:redparS}) can be obtained from the action
(\ref{eq:redSnp}) by putting
\begin{equation}
{\cal N}(t) = \dot{\cal T}(t) \,.  \label{eq:calNT}
\end{equation}
With this replacement, the Hamilton equations of the two actions are
the same, (\ref{eq:motion}). Unlike ${\cal N}(t)$, ${\cal T}(t)$ in
action (\ref{eq:redparS}) can be varied. By varying ${\cal T}(t)$,
we obtain once more the conservation of ${\bf m}(t)$.

     The reduction of our last action, (\ref{eq:Taction}), by the
constraints $\Pi_{T}(r) = 0 = \Pi_{\sf R}(r)$ is trivial. We obtain
\begin{equation}
S \big[ m, p \big] = \int dt \, p \dot{m} \,.  \label{eq:redS}
\end{equation}
We have already observed that the Hamiltonian $h(m,p)$ vanishes, and
that both $m$ and $p$ are constants of motion.

     How are the actions (\ref{eq:redSnp}) and (\ref{eq:redparS})
with a nonvanishing Hamiltonian (\ref{eq:trueh}) related to the
action (\ref{eq:redS})? By a time-dependent canonical
transformation.  Let ${\cal T}(t)$ be a primitive function of ${\cal
N}(t)$, as in (\ref{eq:calNT}). Treat ${\cal T}(t)$ as a prescribed
function of $t$.  Take the function
\begin{equation}
\Omega \big( t, {\bf m}, p \big) = {\bf m} \big( p - {\cal T}(t)
\big) \label{eq:genfun}
\end{equation}
of the old coordinate $\bf m$ and the new momentum $p$, and let it
generate a canonical transformation from $\bf m$ and $\bf p$ to $m$
and $p$:
\begin{equation}
{\bf p} = \partial \Omega \big( t, {\bf m}, p \big) / \partial {\bf
m} = p - {\cal T}(t) \,, \;\;\; m = \partial \Omega \big( t, {\bf
m}, p \big) / \partial p = {\bf m} \,.  \label{eq:timdepcantr}
\end{equation}
Time-dependent canonical transformations change the Hamiltonian:
\begin{equation}
h = {\bf h} + \partial \Omega \big( t, {\bf m}, p \big) / \partial t
= \dot{\cal T}(t) {\bf m} - {\bf m} \dot{\cal T}(t) = 0 \,.
\label{eq:newh}
\end{equation}
Our particular generating function (\ref{eq:genfun}) turns our
particular Hamiltonian ({\ref{eq:trueh}) to zero.

     Equation (\ref{eq:timdepcantr}) reproduces the definition
(\ref{eq:p}) of the momentum $p$. The Hamilton equations
(\ref{eq:motion}) ensure that the new momentum $p$ of equation
(\ref{eq:timdepcantr}) does not change in time. The same conclusion
follows from the new Hamiltonian (\ref{eq:newh}). The three actions
generate the same dynamics.

     The canonical transformation (\ref{eq:timdepcantr}) takes the
canonical momentum ${\bf p}$ at $t$ and transforms it to the value
which it has at an instant when ${\cal T}(t)$ happens to vanish.
The Schwarzschild mass has the same value for any $t$. One can thus
view (\ref{eq:timdepcantr}) as a transformation to `initial data'.

     Similarly, one can view the transition from the unparametrized
action (\ref{eq:Sunpar}) to our final action (\ref{eq:Taction}) as a
time-dependent canonical transformation prior to the reduction.

\section{Quantum Black Holes}

The canonical action (\ref{eq:Taction}) is a good starting point for
the Dirac constraint quantization. The new configuration space is
covered by the coordinates $ T(r), {\sf R}(r)$, and $m$.  The first
two coordinates locate the hypersurface, the third one, $ m$, is the
single degree of freedom of a primordial Schwarzschild black hole.
The constraints are
\begin{equation}
\Pi_{T}(r) = 0\,, \;\;\; \Pi_{\sf R}(r) = 0\,,
\end{equation}
and the Hamiltonian of the system vanishes: $h = 0$.

     The state of the black hole on a hypersurface $T(r), {\sf
R}(r)$ at the label time $t$ should be described by a state
functional $\Psi \big( m ,t;\, T, {\sf R} \big]$ over the
configuration space.  The momenta are represented by the operators
\begin{equation}
\hat{p} = - i\, \partial / \partial m \,, \;\; \hat{\Pi}_{T}(r) = -
i\, \delta / \delta T(r) \,, \;\; \hat{\Pi}_{\sf R}(r) = - i\,
\delta / \delta {\sf R}(r)\,. \label{eq:momop}
\end{equation}
The Dirac rules call for imposing the constraints as operator
restrictions on the state functional:
\begin{equation}
\hat{\Pi}_{T}(r)\,\Psi \big( m ,t;\, T, {\sf R} \big] = 0 \,, \;\;
\hat{\Pi}_{\sf R}(r)\,\Psi \big( m ,t;\, T, {\sf R} \big] = 0 \,.
\label{eq:qconstr}
\end{equation}
Equations (\ref{eq:qconstr}) imply that the state cannot depend on
the embedding variables:
\begin{equation}
\Psi = \Psi ( m, t)\,.  \label{eq:Heisstate}
\end{equation}

     The state must still satisfy the Schr\"{o}dinger equation
\begin{equation}
i\, \dot{\Psi}(m ,t) = \hat{h}\, \Psi ( m,t) \,. \label{eq:Schr}
\end{equation}
Because $\hat{h} = 0$, (\ref{eq:Schr}) ensures that $\Psi$ does not
depend on $t$:
\begin{equation}
\Psi (m, t) = \Psi ( m ) \,.  \label{eq:statehole}
\end{equation}
The state function (\ref{eq:statehole}) describes a superposition of
primordial black holes of different masses. There is not much for it
to do: once prepared, it stays the same on every hypersurface $T(r),
{\sf R}(r)$ and for all $t$.

     Let us compare this description of states with that one which
follows from the unparametrized canonical action (\ref{eq:Sunpar}).
Let us choose the ${\bf \Psi} \big( t; M, {\sf R} \big]$
representation. As before, the $\Pi _{\sf R}(r) = 0$ constraint
implies that $\bf \Psi$ does not depend on ${\sf R}(r)$.  The $M'(r)
= 0$ constraint translates into the statement that $\bf \Psi$ is an
eigenfunction of $\hat{M}(r)$ with a constant eigenvalue $M(r) =
{\bf m}$:
\begin{equation}
\hat{M}'(r)\,{\bf \Psi} = \Big( \hat{M}(r) {\bf \Psi} \Big)' =0 \;\;
\Longrightarrow \;\; \hat{M}(r)\,{\bf \Psi} = {\bf m}\,{\bf \Psi}\,.
\end{equation}
In the $M(r)$ representation
\begin{equation}
{\bf \Psi} \big[M;\,t \big) = \mbox{\boldmath $\psi$} ({\bf m},t)\,
\delta \big( M(r) - {\bf m} \big) \,, \label{eq:redpsi}
\end{equation}
and we can continue working with the coefficient $ \mbox{\boldmath
$\psi$} ({\bf m},t)$.  This coefficient must satisfy the
Schr\"{o}dinger equation with the reduced Hamiltonian
(\ref{eq:trueh}):
\begin{equation}
i\,\dot{ \mbox{\boldmath $\psi$}}({\bf m}, t) = {\cal N}(t) {\bf m}
\; \mbox{\boldmath $\psi$} ({\bf m},t)\,. \label{eq:trueSchr}
\end{equation}
Its solution is
\begin{equation}
 \mbox{\boldmath $\psi$} ({\bf m}, t) = \mbox{\boldmath $\phi$}
({\bf m})\, \exp \big( -i {\bf m} {\cal T}(t) \big)\,,
\label{eq:truestate}
\end{equation}
where ${\cal T}(t)$ is a primitive function to ${\cal N}(t)$, as in
(\ref{eq:calNT}). Unlike (\ref{eq:statehole}), the state function
(\ref{eq:truestate}) oscillates in the $\cal T$ time.

    We have seen that the classical transition from the action
(\ref{eq:Sunpar}) to the action (\ref{eq:Taction}) is achieved by a
time-dependent canonical transformation (\ref{eq:genfun}) -
(\ref{eq:newh}). We want to show that the ensuing quantum theories
are connected by a time-dependent unitary transformation
\begin{equation}
\hat{{\bf W}}(t) = \exp \big(- i {\cal T}(t) \hat{\bf m} \big) \,.
\label{eq:unitary}
\end{equation}

     Let us start in the {\em Heisenberg picture}. The fundamental
Heisenberg operators $\hat{\bf m}(t)$ and $\hat{\bf p}(t)$ of the
unparametrized theory depend on time, and they satisfy the
Heisenberg equations of motion
\begin{equation}
\frac{d \hat{\bf m}(t)}{dt}= \frac{1}{i}\, \big[ \hat{\bf m}(t),
\hat{\bf h}(t) \big]\,, \;\;\; \frac{d \hat{\bf p}(t)}{dt}=
\frac{1}{i}\, \big[ \hat{\bf p}(t), \hat{\bf h}(t) \big]\,,
\end{equation}
with the Heisenberg Hamiltonian $\hat{\bf h}(t)={\bf h}\big(t,
\hat{\bf m}(t), \hat{\bf p}(t)\big)$. The Heisenberg states $|
\mbox{\boldmath $\phi$}_{0} \, \rangle$ refer to $t=0$ and do not
depend on time.

     Change now the fundamental Heisenberg operators $\hat{\bf
m}(t)$ and $\hat{\bf p}(t)$ into new fundamental Heisenberg
operators $\hat{m}(t)$ and $\hat{p}(t)$ by a general time-dependent
unitary operator $\hat{\bf W}(t)= {\bf W} \big( t, \hat{\bf m}(t),
\hat{\bf p}(t) \big)$:
\begin{eqnarray}
\hat{m}(t) \,&\mbox{$:=$}&\, \hat{\bf W}(t)\, \hat{\bf m}(t)\,
\hat{\bf W}^{-1}(t) \,, \label{eq:newopm} \\ \hat{p}(t)
\,&\mbox{$:=$}&\, \hat{\bf W}(t)\, \hat{\bf p}(t)\, \hat{\bf
W}^{-1}(t) \,. \label{eq:newopp}
\end{eqnarray}
It is easy to show that the new Heisenberg operators form a
conjugate pair, and that they satisfy the Heisenberg equations of
motion
\begin{equation}
\frac{d \hat{m}(t)}{dt}= \frac{1}{i}\, \big[ \hat{m}(t), \hat{h}(t)
\big] \,, \;\;\; \frac{d \hat{p}(t)}{dt}= \frac{1}{i}\, \big[
\hat{p}(t), \hat{h}(t) \big] \label{eq:newHeisenbergeq}
\end{equation}
with the new Heisenberg Hamiltonian
\begin{equation}
\hat{h}(t) = \frac{1}{i}\,\frac{\partial \hat{\bf W}(t)}{\partial t}
\, \hat{\bf W}^{-1}(t) + \hat{\bf W}(t)\,\hat{\bf h}(t)\,\hat{\bf
W}^{-1}(t) \,.  \label{eq:newHeisHam}
\end{equation}

     In the {\em Schr\"{o}dinger picture}, the fundamental operators
$\hat{\bf m}$, $\hat{\bf p}$ and $\hat{m}$, $\hat{p}$ become
time-independent, while the state $ | \mbox{\boldmath $\phi$}_{0} \,
\rangle$ is evolved by the respective Hamilton operators:
\begin{eqnarray}
| \mbox{\boldmath $\psi$}(t)\,\rangle & = & {\sf T} \exp \left(
-i\,\int_{0}^{t}dt\, \hat{\bf h}(t) \right) \, | \mbox{\boldmath
$\phi$}_{0}\, \rangle \,, \\ |{\Psi}(t)\, \rangle & = & {\sf T}
\exp\left( -i\,\int_{0}^{t}dt\, \hat{h}(t) \right) \, |
\mbox{\boldmath $\phi$}_{0} \, \rangle \,.
\end{eqnarray}
Here, $\sf T$ stands for the time ordering. We have two
Schr\"{o}\-din\-ger states, $| \mbox{\boldmath $\psi$}(t)\,\rangle$
and $|{\Psi}(t)\,\rangle$, corresponding to two alternative
descriptions, ${\bf m, p, h}(t)$ and $m, p,$$ h(t)$, of the same
quantum system. These states are related by
\begin{equation}
|{\Psi}(t)\,\rangle = {\sf T} \exp \left( -i\,\int_{0}^{t}dt\,
\hat{h}(t) \right)\, {\sf T} \exp \left( i\,\int_{0}^{t}dt\,
\hat{\bf h}(t) \right) \, | \mbox{\boldmath $\psi$}(t)\, \rangle \,.
\label{eq:relatedstates}
\end{equation}

     Apply this general scheme to our simple system. The unitary
operator (\ref{eq:unitary}) yields the Heisenberg fundamental
operators (\ref{eq:newopm}) - (\ref{eq:newopp}),
\begin{eqnarray}
\hat{m}(t)&=&\hat{\bf m}(t)\,,\;\;\;\\ \hat{p}(t)&=&\hat{\bf p}(t) +
{\cal T}(t)\,, \label{eq:meigenval}
\end{eqnarray}
which are related exactly as their classical counterparts
(\ref{eq:timdepcantr}). (For simplicity, we assume that ${\cal
T}(t=0)=0$.) The new Heisenberg Hamilton operator
(\ref{eq:newHeisHam}) vanishes like the classical Hamiltonian
(\ref{eq:newh}). The Heisenberg equations of motion
(\ref{eq:newHeisenbergeq}) then guarantee that the Heisenberg
operators $ \hat{p}(t)$ and $\hat{m}(t)$ are operator constants of
motion. By (\ref{eq:meigenval}), the eigenvalues of the operators
$\hat{\bf m}$ and $\hat{m}$ are the same, ${\bf m}=m$.

     The same situation can be described in the Schr\"{o}dinger
picture. Equation (\ref{eq:relatedstates}) relates the states. In
$m$ - representation,
\begin{equation}
\Psi(m)= \Psi (m, t) = \exp \, \big( i {\cal T}(t){\bf m} \big)\,
\mbox{\boldmath $\psi$} ({\bf m},t) = \mbox{\boldmath $\phi$}({\bf
m})\,.
\end{equation}
This clarifies the relation between the states (\ref{eq:statehole})
and (\ref{eq:truestate}).

     The last of our three actions, (\ref{eq:Scanpi}), has two
additional constraints (\ref{eq:C}). The states now depend on two
more configuration variables $\tau_{\pm}$. The hypersurface
constraints reduce the states to the form $\psi ({\bf m}, \tau_{+},
\tau_{-}, t)$.  The Hamiltonian of the action (\ref{eq:Scanpi})
vanishes, and the Schr\"{o}dinger equation implies that $\psi$
cannot depend on $t$.  The reduced state function must still satisfy
the (reduced) constraints (\ref{eq:C}):
\begin{equation}
\hat{C}_{\pm}\, \psi ({\bf m}, \tau_{+}, \tau_{-}) =0 \;\;
\Longleftrightarrow \;\; \big( \mp i \partial / \partial \tau_{\pm}
+ {\bf m} \big) \psi ( {\bf m}, \tau_{+}, \tau_{-} ) = 0 \,.
\end{equation}
These can be viewed as two Schr\"{o}dinger equations in the proper
times $ \tau_{\pm} $. Their solution is the state function
\begin{equation}
\psi ({\bf m}, \tau_{+}, \tau_{-}) = \phi ({\bf m}) \exp \big( -i\,
{\bf m} (\tau_{+} - \tau_{-} )\, \big) \,.
\end{equation}
This is the same state as (\ref{eq:truestate}), but now written in
terms of the proper times $\tau_{\pm}$ rather than the label time
$t$.  Though they describe it in slightly different ways, our three
actions lead to the same quantum dynamics.

     Because primordial black holes have only one degree of freedom
$m$ which is a constant of motion, their states are rather simple.
Still, there are some interesting questions to ask. The state
function (\ref{eq:statehole}) does not change in time. However, one
can construct significant hypersurface-dependent operators, like the
intrinsic and extrinsic geometry of an embedding $T(r), {\sf R}(r)$,
and ask what their expectation values are. We defer this conceptual
exercise to a later paper.

\section{Inclusion of Sources}

Schwarzschild black holes are empty vessels and what can happen to
them is rather limited. True dynamics requires filling them with
matter. This was the intent of the original BCMN model. Matter can
propagate on the wormhole topology, or it can close the wormhole and
change the topology of $\Sigma$ into ${\rm I\!R}^{3}$. The latter
case is physically more interesting. The following discussion
assumes that the wormhole is closed.

     We must now ask whether what we have done in the vacuum can be
repeated in the presence of matter.

     Introduce a massless scalar field propagating in the spacetime
$ ({\cal M}, \gamma)$:
\begin{equation}
S^{\phi} [\phi ; \gamma] = - \big( 8 \pi \big)^{-1} \int d^{4}X \,
|\gamma|^{1/2}\, \gamma^{\alpha \beta} \phi_{, \alpha} \phi_{,
\beta} \,.
\end{equation}
After the ADM decomposition and midisuperspace reduction by
spherical symmetry, the Lagrangian action takes the form
\begin{equation}
S^{\phi}[\phi;\, \Lambda, R;\, N, N^{r}] = \frac{1}{2} \int dt
\int_{0}^{\infty} dr\, \Big( N^{-1}\,\Lambda R^{2} \big( \dot{\phi}
- N^{r} \phi ' \big)^{2} - N\, \Lambda^{-1}R^{2} \phi'^{2} \Big) \,.
\label{eq:SphiLag}
\end{equation}
By introducing the momentum
\begin{equation}
\pi = \partial L^{\phi} / \partial \dot{\phi} = N^{-1} \Lambda R^{2}
\big( \dot{\phi} - N^{r} \phi' \big) \,,
\end{equation}
we cast the action (\ref{eq:SphiLag}) into canonical form by the
Legendre dual transformation:
\begin{equation}
S^{\phi}[\phi, \pi;\, \Lambda, R;\,N, N^{r}] = \int dt
\int_{0}^{\infty} dr\, \big( \pi \dot{\phi} - NH^{\phi} -
N^{r}H_{r}^{\phi} \big) \,.  \label{eq:Sfield}
\end{equation}
In this process, we obtain the energy density
\begin{equation}
H^{\phi} = \frac{1}{2}\, \Lambda^{-1} \big( R^{-2} \pi^{2} + R^{2}
\phi'^{2} \big) \label{eq:fieldH}
\end{equation}
and momentum density
\begin{equation}
H^{\phi}_{r} = \pi \phi '
\end{equation}
of the scalar field.

     To couple the scalar field to gravity, we add the field action
(\ref{eq:Sfield}) to the gravitational action (\ref{eq:S-ham}). The
variation of the total action with respect to $N$ and $N^{r}$ leads
to Hamiltonian and momentum constraints on the extended phase space:
\begin{equation}
H + H^{\phi} = 0\,, \;\;\; H_{r} + H_{r}^{\phi} = 0 \,.
\label{eq:constmatter}
\end{equation}
Again, up to a point transformation, this is the result obtained by
BCMN \cite{Berger}.

     We arrived at the functional time formalism for the
Schwarzschild black hole by transforming the original geometric
variables into new canonical variables (\ref{eq:can.trans}), and
then into (\ref{eq:canchart}).  Do the charms work in the presence
of sources?

     The message of Section~\ref{sec-MT} is that
(\ref{eq:can.trans}) is a canonical transformation on the geometric
phase space irrespective of any constraints or dynamics. Therefore,
we can introduce the new canonical variables exactly as in the
vacuum spacetime. (The transformation (\ref{eq:canchart}) needs to
be modified, to accommodate the changed topology of $\Sigma$.)

     One cannot, however, expect that this immediately simplifies
the constraints. The underlying physical reason is that the matter
field curves the spacetime in which it propagates, and propagates in
the spacetime which it curves. This is reflected by the presence of
the metric variable $\Lambda$ in the field energy (\ref{eq:fieldH})
in the Hamiltonian constraint (\ref{eq:constmatter}). This variable
is a function (\ref{eq:Lambdanew}) of the new canonical variables.
The Hamiltonian constraint is no longer equivalent to a simple
equation $P_{T}(r) = - M'(r) = 0$, but it provides an implicit
information about how the energy density $M'(r)$ depends on sources.
The structure of this equation is presently being investigated by
Dr.~Joseph Romano in collaboration with the author. An analogous
study of scalar fields coupled to a cylindrical gravitational wave
by Braham \cite{Braham} shows that resolution of such equations is
feasible.

\section{Kruskal Coordinates as Phase-Space Variables}

Our main device has been the reconstruction of the curvature
coordinates $T$ and $R$ from the canonical data.  This turned the
curvature coordinates {\em in spacetime} into canonical coordinates
{\em in phase space}.

     Unfortunately, the Killing time $T$ becomes infinite on the
horizon, and the canonical transformation which leads to it has a
corresponding singularity. From the spacetime picture one knows that
the entire Schwarzschild solution can be covered by a single patch
of spacetime coordinates, the Kruskal coordinates, which are well
behaved on the horizon. It is natural to ask whether these
coordinates, rather than the curvature coordinates, can be
interpreted as canonical coordinates.

     The direct reconstruction of Kruskal coordinates from the
canonical data is cumbersome. It is better to reach them via the
curvature coordinates. Effectively, we are asked to reexpress the
spacetime transformation (\ref{eq:scale}) - (\ref{eq:last}) as a
point transformation on the phase space.

     Because the spacetime transformation involves exponentials, it
is first necessary to turn the curvature coordinates into
dimensionless quantities. This is done by {\em scaling} on the phase
space $\big ( m, T(r), {\sf R}(r);\, p, \Pi_{T}(r), \Pi_{\sf R}(r)
\big)$.  The desired configuration space operation mimics
(\ref{eq:scale}):
\begin{equation}
\bar{T}(r) = \frac{T(r)}{2m}\,, \;\; \bar{\sf R}(r) = \frac{{\sf
R}(r)}{2m} \,, \;\; \bar{m} = m \,.  \label{eq:scaling}
\end{equation}
The scaled curvature coordinates $\bar{T}$ and $\bar{\sf R}$ are
dimensionless, while $\bar{m}$ keeps the dimension of length.

     It is important that the scaling (\ref{eq:scaling}) be done
with the Schwarzschild mass $m$ at infinity rather than with the
mass function $M(r)$. Although these two variables coincide on the
constraint surface, the coordinates $\bar{T}(r)$ and $\bar{\sf
R}(r)$ scaled with the mass function would not have strongly
vanishing Poisson brackets, and hence could not be used as canonical
coordinates on the phase space.

     The configuration transformation (\ref{eq:scaling}) can be
completed into a point transformation on the phase space:
\begin{eqnarray}
\Pi_{T}(r)& = & \frac{\Pi_{\bar{T}}(r)}{2\bar{m}} \,, \\ \Pi_{\sf
R}(r) &=& \frac{\Pi_{\bar{\sf R}}(r)}{2 \bar{m}} \,, \\ p & = &
\bar{p} - \frac{1}{\bar{m}} \int_{-\infty}^{\infty} dr \, \big( \,
\Pi_{\bar{T}}(r) \bar{T}(r) + \Pi_{\bar{\sf R}}(r) \bar{\sf R}(r) \,
\big) \,.
 \end{eqnarray}
Inversely,
\begin{eqnarray}
\Pi_{\bar{T}}(r)& = & 2m \Pi_{\bar{T}}(r) \,, \\ \Pi_{\bar{\sf
R}}(r)&=& \; 2m \Pi_{\sf R}(r) \,, \\ \bar{p} & = & p + \frac{1}{m}
\int_{-\infty}^{\infty} dr \, \big( \, \Pi_{T}(r) T(r) + \Pi_{{\sf
R}}(r) {\sf R}(r) \, \big) \,.
 \end{eqnarray}
{}From here, we can figure out the dimensions of the momenta. The
unscaled coordinates $T$ and $\sf R$ have the dimension of length,
while the conjugate momenta $P_{T}$ and $P_{\sf R}$ are
dimensionless.  The scaling reverts these dimensions: The scaled
coordinates $\bar{T}$ and $\bar{\sf R}$ are dimensionless, while the
scaled momenta $\Pi_{\bar{T}}$ and $\Pi_{\bar{\sf R}}$ have the
dimension of length.  Scaling does not change the dimension of the
discrete variables: $m$ and $\bar{m}$, and $p$ and $\bar{p}$ all
have the dimension of length.

     Because the transformation from curvature coordinates to
Kruskal coordinates is double-valued, it is better to write the
transformation {\em from} Kruskal coordinates {\em to} curvature
coordinates. The configuration part of this transformation follows
the pattern of (\ref{eq:Kruskal-curv}):
\begin{equation}
\bar{T}(r)=\ln|V(r)| - \ln |U(r)| \,,\;\;\; \bar{\sf R}(r)={\cal
R}\big(U(r)V(r)\big)\,. \label{eq:Kruskal-conf}
\end{equation}

     By differentiating (\ref{eq:last}) with respect to $\bar{T}$
and $\bar{R}$ we obtain the Jacobi matrix
\begin{eqnarray}
U_{,\bar{T}}&=& -\frac{1}{2} U\,, \;\;\;\; U_{,\bar{R}}=
\frac{1}{2}UF^{-1}(\bar{R})\,, \vspace{2 mm} \\ V_{,\bar{T}}&=&
\;\;\;\frac{1}{2} V\,, \;\;\;\; V_{,\bar{R}}=
\frac{1}{2}VF^{-1}(\bar{R})\,. \label{eq:Jacobi}
\end{eqnarray}
We took into account that
\begin{equation}
\frac{d \bar{R}^{\ast}\big(\bar{R})}{d \bar{R}}=
F^{-1}\big(\bar{R}\big) = \big(1-\bar{R}^{-1}\big)^{-1}.
\end{equation}
The completion of (\ref{eq:Kruskal-conf}) into a point
transformation is straightforward:
\begin{eqnarray}
P_{\bar{T}}(r)& = & V_{,\bar{T}}(r) \Pi_{V}(r) + U_{,\bar{T}}(r)
\Pi_{U}(r) \nonumber \\ {} & = & \frac{1}{2} \Big( V(r) \Pi_{V}(r) -
U(r) \Pi_{U}(r) \Big) \,, \label{eq:KruskalmomentaT} \vspace{3 mm}
\\ P_{\bar{\sf R}}(r)& = & V_{,\bar{R}}(r) \Pi_{V}(r) +
U_{,\bar{R}}(r) \Pi_{U}(r) \nonumber \\ {} &=& F^{-1}\Big( {\cal
R}\big(U(r)V(r)\big)\Big)\, \frac{1}{2} \Big( V(r) \Pi_{V}(r) + U(r)
\Pi_{U}(r) \Big)\,.  \label{eq:KruskalmomentaR}
\end{eqnarray}

     We assume that the Kruskal variables $U(r), \Pi_{U}(r), V(r),
\Pi_{V}(r)$ are continuous. The transformation equations
(\ref{eq:Kruskal-conf}) and (\ref{eq:KruskalmomentaT}) -
(\ref{eq:KruskalmomentaR}) reveal that the curvature variables
$\bar{\sf R}(r)$ and $\Pi_{\bar{T}}(r)$ will also be continuous,
whereas $\bar{T}(r)$ and $\Pi_{\bar{\sf R}}(r)$ become infinite when
$U(r)=0$ or $V(r)=0$.

\section{Acknowledgements}

     I want to thank the participants of the ITP workshop on Small
Scale Structure of Spacetime, and many other people with whom I
discussed this work during its gestation. Several talks with Charles
Torre helped me to clarify my ideas about parametrized infinities
and canonical charts. I am especially grateful to Joseph Romano for
his critical reading of my notes and corrections of my slips.

     This research was supported in part by the NSF grants
PHY89-04035 and PHY-9207225, and by the U.S.-Czech Science and
Technology Grant No~92067.

 \end{document}